\documentclass[a4paper,fleqn,usenatbib]{mnras}

\usepackage{mathptmx}

\usepackage[T1]{fontenc}
\usepackage{ae,aecompl}

\usepackage{graphicx}	
\usepackage{amsmath}	
\usepackage{amssymb}	
\usepackage{textcomp}
\usepackage{color}
\usepackage{multicol}
\usepackage{longtable}
\usepackage{rotating}
\usepackage{pdflscape}
\usepackage{natbib}
\usepackage{enumitem}
\usepackage{appendix}
\usepackage{url}
\usepackage{float}
\usepackage{tabularx}
\usepackage{booktabs}
\usepackage[caption = false]{subfig}
\usepackage{scalerel}

\title[$H(z)$ from only BCGs]{An independent estimate of $H(z)$ at $z=0.5$ from the stellar ages of brightest cluster galaxies\thanks{based on observations made with the South African Large Telescope (SALT).}}

\author[Loubser et al.]{S. Ilani Loubser$^{1,2}$\thanks{E-mail:Ilani.Loubser@nwu.ac.za (SIL)}, Adebusola B. Alabi$^{1}$, Matt Hilton$^{3,4}$,  Yin-Zhe Ma$^{5}$, Xin Tang$^{6}$,   
\newauthor Narges Hatamkhani$^{7}$, Catherine Cress$^{8}$, Rosalind E. Skelton$^{7}$ and S. Andile Nkosi$^{1}$ \\
$^{1}$Centre for Space Research, North-West University, Potchefstroom 2520, South Africa\\
$^{2}$National Institute for Theoretical and Computational Sciences (NITheCS), Potchefstroom 2520, South Africa\\
$^{3}$Wits Centre for Astrophysics, School of Physics, University of the Witwatersrand, Private Bag 3, Johannesburg 2050, South Africa\\
$^{4}$School of Mathematics, Statistics \& Computer Science, University of KwaZulu-Natal, Westville Campus, Durban 4041, South Africa\\
$^{5}$Department of Physics, Stellenbosch University, Matieland 7602, South Africa\\
$^{6}$Department of Physics and Astronomy, University of Sussex, Falmer Campus, Brighton, BN1 9QH, UK \\
$^{7}$South African Astronomical Observatory, PO Box 9, 1 Observatory Road, Observatory, Cape Town 7925, South Africa \\
$^{8}$Department of Mathematical Sciences, University of South Africa, Florida Park, Roodepoort 1709, South Africa \\
}

\date{Accepted 2025 June 04. Received 2025 April 23; in original form 2025 January 27.}

\pubyear{2025}

\begin{document}
\label{firstpage}
\pagerange{\pageref{firstpage}--\pageref{lastpage}}
\maketitle

\begin{abstract}
Several cosmological observations (e.g., Cosmic Microwave Background (CMB), Supernovae Type Ia, and local distance ladder measurements such as Cepheids) have been used to measure the global expansion rate of the Universe, i.e., the Hubble constant, $H_{0}$. However, these precision measurements have revealed tensions between different probes that are proving difficult to solve. Independent, robust techniques must be exploited to validate results or mitigate systematic effects. We use the Cosmic Chronometer (CC) method, which leverages the differential age evolution of passive galaxies, to measure $H(z)$, without any assumption of the underlying cosmology. Unlike previous CC studies, we used only brightest cluster galaxies (BCGs), the oldest and most massive galaxies in the Universe, to construct a pure and homogeneous sample. In this work we used a sample of 53 BCGs in massive, Sunyaev-Zel'dovich selected galaxy clusters (0.3 $< z <$ 0.7) with Southern African Large Telescope (SALT) spectroscopic observations. We used optical spectra to measure D4000$_{\rm n}$ of the BCGs to obtain a new direct measurement of $H(z) = 72.1 \pm 33.9(\rm stat) \pm 7.3$(syst) km s$^{-1}$ Mpc$^{-1}$ at $z=0.5$. By using BCGs, we significantly reduced the systematic errors to 10\% by minimising the stellar mass and metallicity dependence of the method. The dominant uncertainty, and limitation for our study, is statistical, and we need larger, homogeneous samples of the oldest, most massive galaxies. By using the $Planck$+BAO prior of $\Omega_{m}$ and $\Omega_{\Lambda}$, the projected Hubble constant is $H_{0}$ =  $54.6 \pm 25.7(\rm stat) \pm 5.5$(syst) km s$^{-1}$ Mpc$^{-1}$, consistent with both CMB and Cepheid measurements.
\end{abstract}




\section{Introduction}
\label{Section:introduction}


Several cosmological probes, e.g., Cosmic Microwave Background (CMB, such as the $Wilkinson\ Microwave\ Anisotropy\ Probe$ \citep{Hinshaw2013} and $Planck$ survey \citep{Aghanim2020}), Supernovae Type Ia (SNIa, \citealt{Abbott2024}), and Baryon Acoustic Oscillations (BAO) from spectroscopic galaxy surveys (such as eBOSS, \citealt{Adame2024}) have been used to better understand the nature of the accelerated expansion of the Universe. However, different measurements give different values regarding the Hubble constant ($H_{0}$), one of the most fundamental parameters in the Universe \citep{Verde2019, Riess2022}. In particular, the $H_{0}$ constrained by the CMB angular power spectra from different surveys give values ($H_{0} = 67.4 \pm 0.5$ km s$^{-1}$ Mpc$^{-1}$ from \citealt{Aghanim2020}; and $H_{0} = 67.6 \pm 1.1$ km s$^{-1}$ Mpc$^{-1}$ from \citealt{Aiola2020}), that are 4--5$\sigma$ lower than the measured $H_{0}$ value from the Cepheid distance ladder ($H_{0} = 73.04 \pm 1.04$ km s$^{-1}$ Mpc$^{-1}$ from \citealt{Riess2022}). Other observations, such as strong gravitational lensing \citep{Wong2024}, quasars \citep{Risaliti2015}, gamma-ray bursts \citep{Amati2008}, and gravitational waves as standard sirens \citep{Schutz1986, Abbott2017-GW}, tend to give values that sit somewhere in between. Before claiming the necessity of invoking new physics to explain this discrepancy, it is necessary to develop independent and complementary methods to test the robustness of existing measurements and give hints of the possible systematics \citep{Albrecht2006, diValentino2021, Moresco2022, Jimenez2023, Bergamini2024}.   

In the standard Friedmann–Lemaitre–Robertson–Walker (FLRW) spacetime metric, the Hubble parameter as a function of redshift, $H(z)$, is related to the differential ageing of the Universe ($dt_{U}$) via:
\begin{equation}
\label{eqn:Hz}
H(z) = - \frac{1}{1+z} \frac{dz}{dt_{U}}. 
\end{equation}
Therefore, by measuring the age difference between two passively evolving galaxies that formed at the same time but are separated by a small redshift interval, one can calculate the derivative $dz/dt_{U}$, which can determine the Hubble parameter at that redshift. The method of using a homogeneous population of massive, passively evolving galaxies to trace $dt_{U}$, is called Cosmic Chronometers (CC, \citealt{Jimenez2002}). The CC method has been successfully applied to measure $H(z)$ up to $z \sim 2$ \citep{Stern2010, Moresco2012, Zhang2014, Moresco2016, Ratsimbazafy2017, Borghi2022a, Borghi2022b, Tomasetti2023}. Between redshifts of $z = 1 \sim 2$, differential age measurement can also be applied to constrain the equation of state of dark energy \citep{Jimenez2002, Jimenez2003, Simon2005}. The advantage of CC compared to other probes is that it can provide a direct estimate of $H(z)$ without any assumption of the underlying cosmology (beyond that of the FLRW metric in Eq.\ \ref{eqn:Hz}). 

The best targets for CC are the oldest and most massive galaxies. There is general agreement that these massive ($M_{\star} > 10^{11} M_{\sun}$) elliptical galaxies formed more than 90\% of their stellar mass very rapidly (within 0.3 Gyr) at redshifts $z >$ 2 -- 3. They only experienced minor subsequent episodes of star formation, making them the oldest objects at all redshifts (e.g., \citealt{Cimatti2004, Treu2005, Renzini2006, Thomas2010}). Since CC is based on differential ages, rather than absolute age determination for the stellar populations of the galaxies, and the differential ages are estimated in thin redshift slices, it minimises the potential effects of uncertainties in galaxy evolution over large redshift ranges \citep{Moresco2012}. Similarly, the small spread in age among stars in an elliptical galaxy does not affect CC results because the method relies only on the shift in the average stellar age as a function of redshift. In the CC method, the ages of massive, red, passive galaxies are determined with stellar population models (\citealt{Jimenez2003, Simon2005, Stern2010}, see also the discussion in \citealt{Kjerrgren2023}), either using combinations of Lick indices \citep{Borghi2022a, Borghi2022b}, the age-sensitive spectral break at 4000 \AA{} rest frame (hereafter D4000$_{\rm n}$, \citealt{Moresco2011, Moresco2012, Moresco2016}), or full-spectrum fitting \citep{Zhang2014, Ratsimbazafy2017} or both indices and full-spectrum fitting \citep{Jiao2023}. Photometry, trained on spectroscopic samples, can also be used \citep{Tomasetti2023}, or in combination with spectroscopy. 
 

According to the downsizing scenario, galaxy mass is the main driver of galaxy formation and evolution, with more massive galaxies forming their stars at earlier cosmic epochs compared to less massive ones \citep{Thomas2010}. For this reason, multiple parallel age--redshift relations for different masses of populations are expected. When we consider all galaxies, regardless of mass, they form an envelope in the age--redshift plane. Large spectroscopic samples with very high number statistics are then needed to fill this envelope to probe the upper edge (the evolution of the oldest, most massive galaxies). The slope of this upper edge of the envelope directly relates to $H(z)$ (equation \ref{eqn:Hz}), and since massive, red galaxies evolve slowly at lower redshifts, the slope is shallow by definition and $H(z)$ is extremely sensitive to uncertainties in the masses of galaxies observed at different redshifts. Even when only massive, passively evolving galaxies are used, they form a broad part of this envelope (see e.g., fig. 4 (right) in \citealt{Borghi2022b}, and fig. 5 in \citealt{Jiao2023}), and there is a danger of mixing galaxy populations which would directly have a significant effect on the value of $H(z)$. 

Another effect is the progenitor bias, where the progenitors of younger elliptical galaxies can be missing at higher redshifts due to observational constraints \citep{Bender1998}. This may lead to a flattening of the age-redshift relation, if measured over long redshift ranges, and thus to a bias in $H(z)$. It is important to investigate whether such systematics can be eliminated by selecting the most homogeneous population of the oldest, massive galaxies, i.e., only brightest cluster galaxies (BCGs) in thin redshift slices -- even if this results in a limitation of the sample size. Although other CC studies have included small numbers of BCGs in their samples, here we will only use BCGs between 0.3 $< z <$ 0.7, and only from the BEAMS (BCG evolution with ACT, MeerKAT and SALT) sample \citep{Hilton2021}, described in detail in Section \ref{data}. 

On the other hand, the galaxies that belong to the overdense region of a cluster might not be representative of the rest of the Universe. In comparing clusters at different redshifts, one must be sure that the clusters have similar general properties (e.g., belong to the same richness class) and that differences between the clusters that are being compared do not introduce environmental biases \citep{Jimenez2002}. Sunyaev-Zel'dovich (SZ) cluster samples are approximately mass selected, and a representative sample of the most massive clusters in the Universe. It is also essential to only use massive galaxies that are truly passive, i.e., without a younger stellar component contaminating the differential age measurement \citep{Lopez2018}. We perform various checks for the presence of star formation or Active Galactic Nuclei (AGN) in our sample of BCGs. 


Our main goal is to make a CC measurement at $z = 0.5$ using a homogeneous population of the most massive galaxies (BCGs in massive clusters). The $H(z)$ measurement in this work does not rely on the assumption of a cosmological model, but only on the minimal assumption of the FLRW metric. For the projection of $H(z=0.5)$ to $H_{0}$, we adopt a standard $\Lambda$CDM cosmology model with the $Planck$+BAO prior of $\Omega_{m}$ and $\Omega_{\Lambda}$ values. The rest of the paper is organised as follows. In Section \ref{data}, we describe the cluster and BCG sample selection, the observations, data reduction, BCG selection from the long-slit observations and redshift determination, and the stacking of the optical spectra. In Section \ref{measurements}, we describe our D4000$_{\rm n}$ measurements in detail, and in Section \ref{calibration}, we calibrate the D4000$_{\rm n} - z$ relation to obtain $H(z)$ at $z=0.5$. In Section \ref{systematics}, we also assess all systematic effects, and compare to previous CC measurements (Section \ref{previousCC}). We summarise and conclude in Section \ref{summary}.  

\section{Data}
\label{data}

\subsection{ACT clusters, and the BEAMS sample of BCGs}

The galaxies used in this study are drawn from the BEAMS\footnote{\url{https://astro.ukzn.ac.za/~beams/}} project, a Large Science Programme (2019-1-LSP-001 and 2022-2-MLT-003, PI: Hilton) on the Southern African Large Telescope (SALT) that aimed to obtain long-slit spectroscopic observations of 100--150 central cluster galaxies in a representative sample of SZ-selected clusters detected by the Atacama Cosmology Telescope (ACT; \citealt{Swetz2011}). The selection of BEAMS targets was drawn from a pool of 172 clusters at 0.3 $< z <$ 0.8, detected with ACT SZ with signal-to-noise ratio (S/N) > 5 in a preliminary cluster catalogue based on the ACT DR4 maps \citep{Aiola2020} and located within the Dark Energy Survey (DES; \citealt{Abbott2018}) footprint (in order to ensure good optical photometric data were available). SALT data were obtained for 113/172 possible targets. After the BEAMS target list was defined, deeper ACT data were used to make the final ACT DR5 cluster catalogue \citep{Hilton2021}, and 171/172 of the BEAMS targets are found in the final ACT DR5 catalogue. The richness-based weak-lensing mass estimates (see \citealt{Hilton2021}) of the 113 observed BEAMS clusters span the range 2.4 < $M_{\rm 500c}^{\rm Cal}$ < 19 $\times$ 10$^{14}$\,M$_{\sun}$ (median $M_{\rm 500c}^{\rm Cal} = 4.8 \times 10^{14}$\,M$_{\sun}$), and represent an approximately mass-limited cluster sample in the redshift range 0.3 $< z <$ 0.8 due to the SZ-selection.

\subsection{SALT observations and data reduction} 

BEAMS obtained Robert Stobie Spectrograph (RSS) long-slit spectroscopy with resolution $\sim 3$ \AA{} and rest-frame wavelength coverage 3800 – 5100 \AA{} (to cover the range from D4000$_{\rm n}$ to [OIII] 5007 \AA{}). We used the PG1300 grating in combination with a 1.25\arcsec\ wide slit for BCGs at $z <$ 0.65. At $z >$ 0.65, we used the PG0900 grating to ensure that we covered the needed wavelength range. The exposure times were $3\times800$ seconds for $z <$ 0.65, and double that for $z >$ 0.65. We also observed arcs for wavelength calibration directly after each set of science exposures and spectrophotometric standard stars for relative flux calibration.  

The SALT data have been processed with a modified version of the \texttt{RSSMOSPipeline} package\footnote{\url{https://github.com/mattyowl/RSSMOSPipeline}} described in \citet{Hilton2018}. Basic corrections and calibrations such as overscan, gain, cross-talk corrections, and mosaicking are performed by the SALT science pipeline, $\mathtt{PySALT}$\footnote{\url{https://pysalt.salt.ac.za}} \citep{Crawford2010}, developed in the Python/PyRAF environment. We perform sky subtraction by applying the optimal background subtraction algorithm introduced in \citet{Kelson2003}. For each 2D sky+source slitlet from the \texttt{RSSMOSPipeline}, we obtained a 2D sky model using CCD rows far away from the source, which we over-sample at the sub-pixel level (typically by a factor of 20). We then fit cubic splines to these oversampled sky rows and reproject to the entire 2D slitlet grid to construct the sky model. Over-sampling the background ensures that sharp gradients from the bright sky-lines are properly modelled and adequately removed from the sky-subtracted frame. 

The 8\arcmin\ long-slit was positioned to observe more than one galaxy, and we needed to identify which of the extracted one-dimensional spectra (from \texttt{RSSMOSPipeline}) contained the BCG. Identification of the BCG was performed via a visual inspection of the DES imaging with the aid of finder charts from the SALT observations. The redshifts were measured using the \texttt{XCSAO} task of the \texttt{RVSAO IRAF} package \citep{Kurtz1998} and verified by visual inspection. We consider redshifts measured from spectra in which two or more strongly detected features were identified (for example, the H and K lines due to Ca II) to be secure. Of the 113 BCGs, 14 do not have reliable redshifts. We also compared the redshift measured for the BCG with the redshift of the cluster \citep{Hilton2021}, and removed 3 BCGs for which there were discrepancies between the BCG and cluster redshifts. Of the remaining 96 BCGs, 18 BCGs are at $z>0.7$ and very faint. We restricted ourselves to the redshift range of $0.3 < z < 0.7$ to use spectra with sufficient S/N. 

\subsection{Stacking spectra}

To measure the differential stellar population properties, data with reliable S/N are required. We stack the data in small redshift bins, similar to \citet{Moresco2012, Moresco2016, Borghi2022b}. We stack BCGs that can be stacked together without a single redshift bin ($\Delta z$) becoming larger than $\sim$ 0.02. The redshift bin width of $\Delta z < 0.02$ corresponds to a cosmic time at $z = 0.5$ for which the D4000$_{\rm n}$ index is expected to change by $< 0.008$ (the exact value depends on the stellar population model, metallicity, etc. as discussed in Section \ref{calibration}). This expected change in D4000$_{\rm n}$ across the redshift bin is at least an order of magnitude smaller than our measurement error on D4000$_{\rm n}$. The redshift bins that we use are shown in Table \ref{tableLick}. 

\begin{table*}
\caption{Redshifts and index measurements for the stacked spectra.}    
\label{tableLick}      
\centering                         
\begin{tabular}{c c c c c c c} 
\hline
Stack & Median $z$ & $\Delta z$ & Ca II H/K & D4000$_{\rm n}$ & $\#$ BCGs in Stack & Velocity dispersion  \\
 &  &  &  &  &  &  (km s$^{-1}$) \\
\hline 
1 & 0.3300 $\pm$ 0.0044 & 0.014 & 1.175 $\pm$ 0.456 & 2.207 $\pm$ 0.067 & 4 &  301 $\pm$ 41   \\
2 & 0.3565 $\pm$ 0.0041 & 0.020 & 0.727 $\pm$ 0.142 & 2.332 $\pm$ 0.055 & 6 & 347 $\pm$ 36  \\
3 & 0.3795 $\pm$ 0.0038 & 0.020 & 0.982 $\pm$ 0.324 & 1.994 $\pm$ 0.051 & 6 &  261 $\pm$ 51 \\
4 & 0.3960 $\pm$ 0.0044 & 0.021 & 0.924 $\pm$ 0.292 & 2.069 $\pm$ 0.072 & 5 &  276 $\pm$ 59  \\
5 & 0.4275 $\pm$ 0.0061 & 0.020 & 1.033 $\pm$ 0.265 & 1.951 $\pm$ 0.051 & 4 &   276 $\pm$ 26  \\
6 & 0.4500 $\pm$ 0.0041	& 0.020 & 0.700 $\pm$ 0.174 & 2.019 $\pm$ 0.091 & 6 &  270 $\pm$ 64  \\
7 & 0.4920 $\pm$ 0.0072 & 0.020 & 1.028 $\pm$ 0.517 & 2.026 $\pm$ 0.064 & 3 &   194 $\pm$ 39  \\
8 & 0.5380 $\pm$ 0.0037 & 0.016 & 1.222 $\pm$ 1.587 & 2.051 $\pm$ 0.086 & 4 &   502 $\pm$ 121 \\
9 & 0.5575 $\pm$ 0.0060 & 0.018 & 1.044 $\pm$ 0.488 & 2.013 $\pm$ 0.126 & 4 &   324 $\pm$ 57  \\
10 & 0.6055 $\pm$ 0.0056 & 0.019 & 0.691 $\pm$ 0.376 & 2.150 $\pm$ 0.089 & 4 &   315 $\pm$ 70  \\
11 & 0.6200 $\pm$ 0.0030 & 0.011 & 0.506 $\pm$ 0.163 & 1.964 $\pm$ 0.082 & 4 &   230 $\pm$ 70  \\
12 & 0.6500 $\pm$ 0.0080 & 0.020  & 0.582 $\pm$ 0.384  & 2.106 $\pm$ 0.096  &  3 &  339 $\pm$ 63  \\
\hline                                   
\end{tabular}
\end{table*}	

During stacking, we also visually inspect every spectrum, including checking for any emission lines signifying recent star formation. Although, in general, typically only a modest fraction ($\sim$ 15\%) of BCGs show strong optical line emission, this fraction becomes larger ($\sim$ 71\%) for BGCs in cooling flow clusters \citep{Edwards2007}, and even greater ($\sim$ 96\%) in local cooling flow clusters with small offsets ($\sim$ 15 kpc) between their X-ray emission and the BCG \citep{Sanderson2009}. We do not stack any BCGs with skyline residuals or BCGs where there is any indication of possible emission lines. Overall, less than $\sim 5\%$ of the full BEAMS sample contain clear emission lines in their spectra. As a further check, we fit for any residual emission lines in the stacked spectra in Section \ref{fitting}, and we also measure the Ca H and K line ratios as additional confirmation that the galaxies are passive in Section \ref{Cal}. 

\begin{figure}
\centering
\includegraphics[scale=0.5, trim= 50 250 40 270, clip]{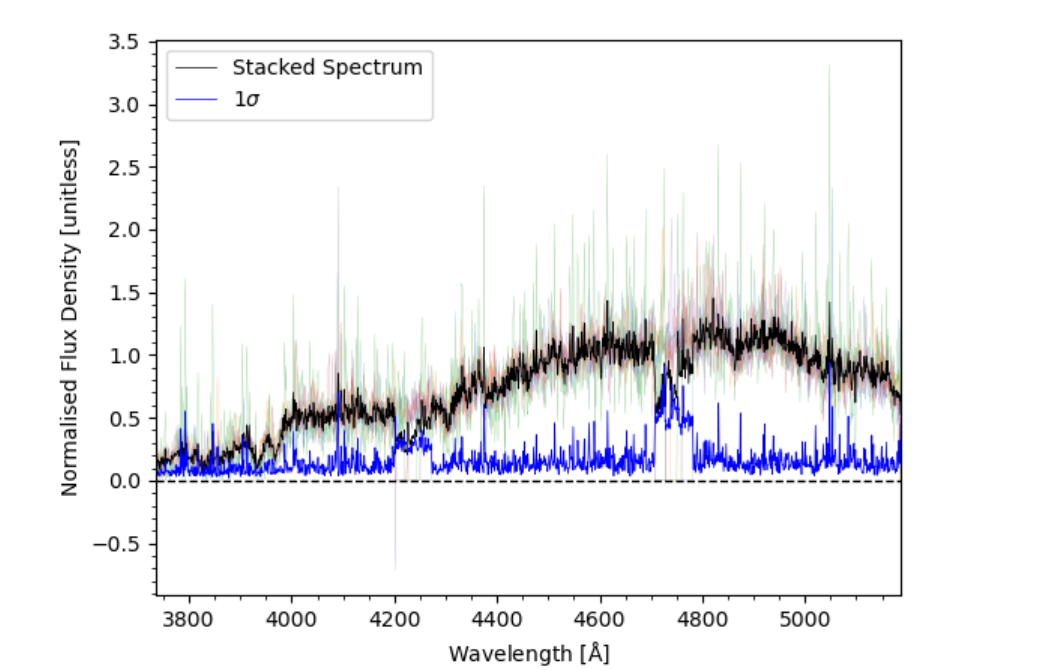}
   \caption{A $\mathtt{specstack}$ stacked spectrum of five BCGs at $z=0.36$ for illustration. The black line is the resultant stacked spectrum, and the blue indicate the 1$\sigma$ error. The coloured lines in the background show the five individual, lower S/N, spectra. The SALT chip gaps can be seen at approximately 1/3 and 2/3 of the wavelength range, and these regions are masked during measurements or full-spectrum fitting.}
\label{fig:stacked_spectraD3}
\end{figure}

We used $\mathtt{specstack}$\footnote{\url{https://specstack.readthedocs.io/en/latest/}} to stack the BCG spectra in each redshift bin into a single averaged spectrum \citep{Thomas2019}. We de-redshift the wavelength grid to $z = 0$, and normalise the flux to the region between 4500 to 4600 \AA{} for each BCG spectrum. This does not change the shape of the continuum spectra; it merely normalises all individual spectra to have a flux density of 1 (unitless) to stack them with equal weight. The $\mathtt{specstack}$ routine determines the maximal wavelength window in which the stacked spectrum can be computed with all individual spectra to ensure that each point in the stacked spectrum is computed with the same amount of individual flux points. The final stacked spectrum is the mean of all the individual spectra after a sigma-clipping algorithm is applied. The binning in wavelength increments of the new stacked spectrum is also $\sim 3$ \AA{}. We show an example of a stacked spectrum in Figure \ref{fig:stacked_spectraD3}. 

From the 78 BCGs, we finally stack 53 BCGs. The other spectra were not stacked, mostly because the first chip gap in the SALT spectra was too close to the D4000$_{\rm n}$ measurement (see Figure \ref{fig:stacked_spectraD3}), or the redshift of the BCG was not close enough to other BCGs to be stacked while keeping $\Delta z$ less than 0.02, or possible emission lines or poor quality of the spectrum. From the 53 BCGs, we create 12 stacked spectra, shown in Figure \ref{fig:stacked_spectraD4000n}. We list the BCGs in the Appendix \ref{bcgs_table}. The mass estimates (see \citealt{Hilton2021}) of the 53 BEAMS clusters used here span the range 2.66 < $M_{\rm 500c}^{\rm Cal}$ < 10.98 $\times$ 10$^{14}$\,M$_{\sun}$ (median $M_{\rm 500c}^{\rm Cal} = 5.37 \times 10^{14}$\,M$_{\sun}$). 

\section{Measurements}
\label{measurements}

\subsection{Full-spectrum fitting}
\label{fitting}

We use the Penalized PiXel-Fitting method\footnote{\url{https://pypi.org/project/ppxf/}} (\texttt{pPXF}, \citealt{Cappellari2004, Cappellari2017, Cappellari2023}) to fit the stacked spectra with the E-MILES simple stellar populations (SSPs) templates \citep{Vazdekis2015}. We simultaneously fit the stellar features for kinematics and stellar populations and include emission lines (forbidden and Balmer) to detect if any emission lines are present over our full wavelength range.  We corrected the velocity dispersion for the instrumental resolution as well as the resolution of the spectral templates to obtain the intrinsic broadening of the stacked spectra. We do not use any penalisation (bias) and do not add an additive polynomial. We also do not include gas reddening but instead use a multiplicative polynomial. We fit for an estimation of the velocity dispersion, and simultaneously confirm that the velocity is near zero (the redshift of the stacked spectrum), that there are no emission lines present, and measure the stellar population (SSP-equivalent) age and metallicity. The fits, together with the age and metallicity estimates, are shown in Appendix \ref{ppxf_fits} and the velocity dispersion in Table \ref{tableLick}. 

We lose a significant part of the full spectrum as a result of the multiple CCD chip gaps, as well as losing short wavelength ranges at the beginning and end of the stacked spectrum, as can be seen in Figures \ref{fig:stacked_spectraD3} and \ref{fig:stacked_spectra_ppxf} (the light blue bands). As a result, the uncertainty on the velocity dispersion is large, especially for stack 8 for which we obtained a particularly poor fit. For our purpose, it is sufficient to confirm that the velocity dispersion, age and metallicity are at the upper end of the range for the most massive, old, and metal-rich galaxies. Using different combinations of models and libraries can provide information on the systematic uncertainty involved in the fit of the model \citep{Groenewald2014, Loubser2016, Loubser2021, Borghi2022b}. However, because of our limited wavelength coverage and the fact that we cannot accurately separate the various systematic errors and degeneracies related to the model fitting (e.g., choice of library, models, initial mass function (IMF), assumed star formation histories), we opt to use direct index measurements. The selection of a smaller, specific wavelength region in the spectrum increases statistical errors on the measurements since less information is used, but it avoids some of the systematics related to assumptions made in full-spectrum fitting. 

\subsection{Direct observables: Indices}

We use $\mathtt{PyLick}$\footnote{\url{https://gitlab.com/mmoresco/pylick/}.} to measure Ca K, Ca H (+ H$\epsilon$), and D4000$_{\rm n}$ (see Table \ref{tableLick}). Since we use D4000$_{\rm n}$ which is just the flux ratio in two bands, and the ratio of the adjacent features Ca H (+ H$\epsilon$) and Ca K, it is not necessary to correct the ratios for dispersion. We also test whether the wavelength range choice for normalisation during stacking affects the index measurements and find that the measurements do not depend on the wavelength range chosen.

\begin{figure*}
\captionsetup[subfloat]{farskip=-0pt,captionskip=-0pt}
\centering
\subfloat{\includegraphics[scale=0.28]{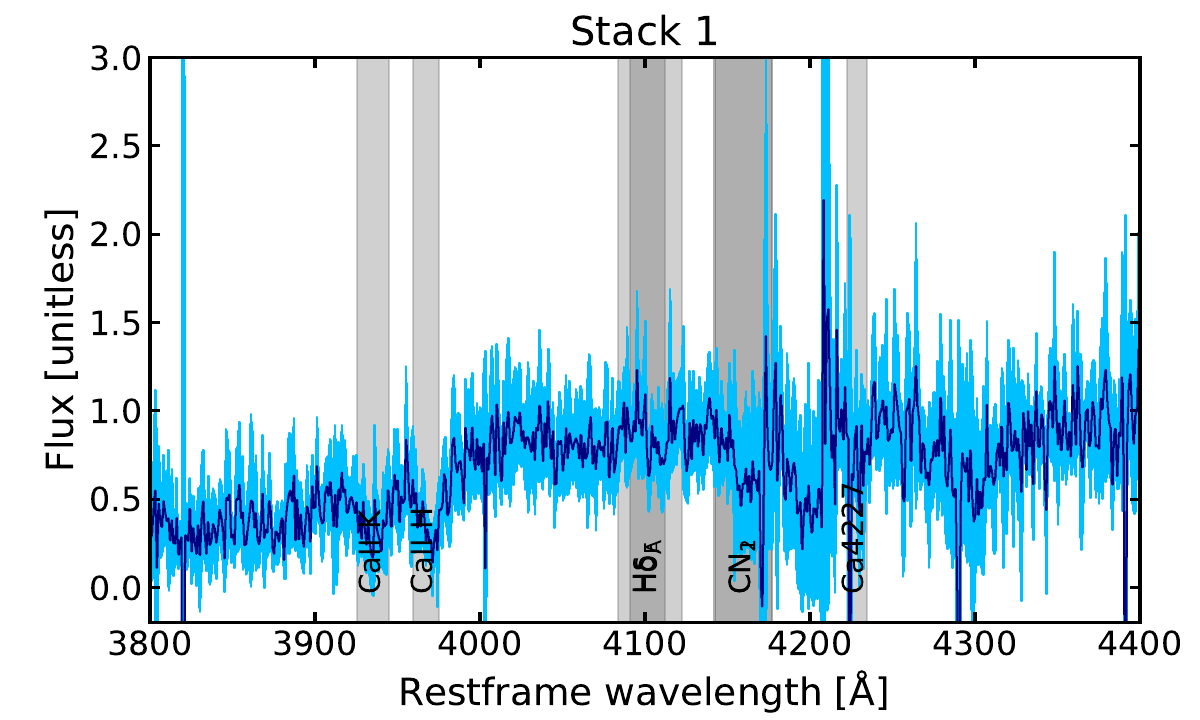}}
\subfloat{\includegraphics[scale=0.28]{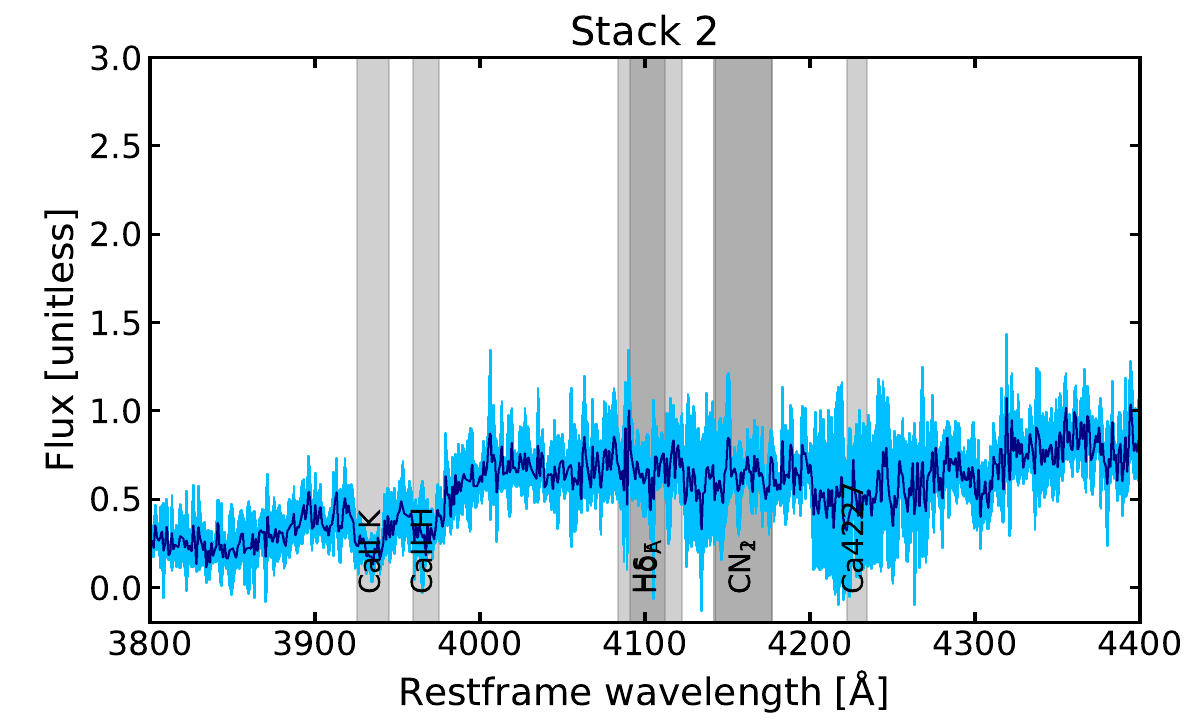}}
\subfloat{\includegraphics[scale=0.28]{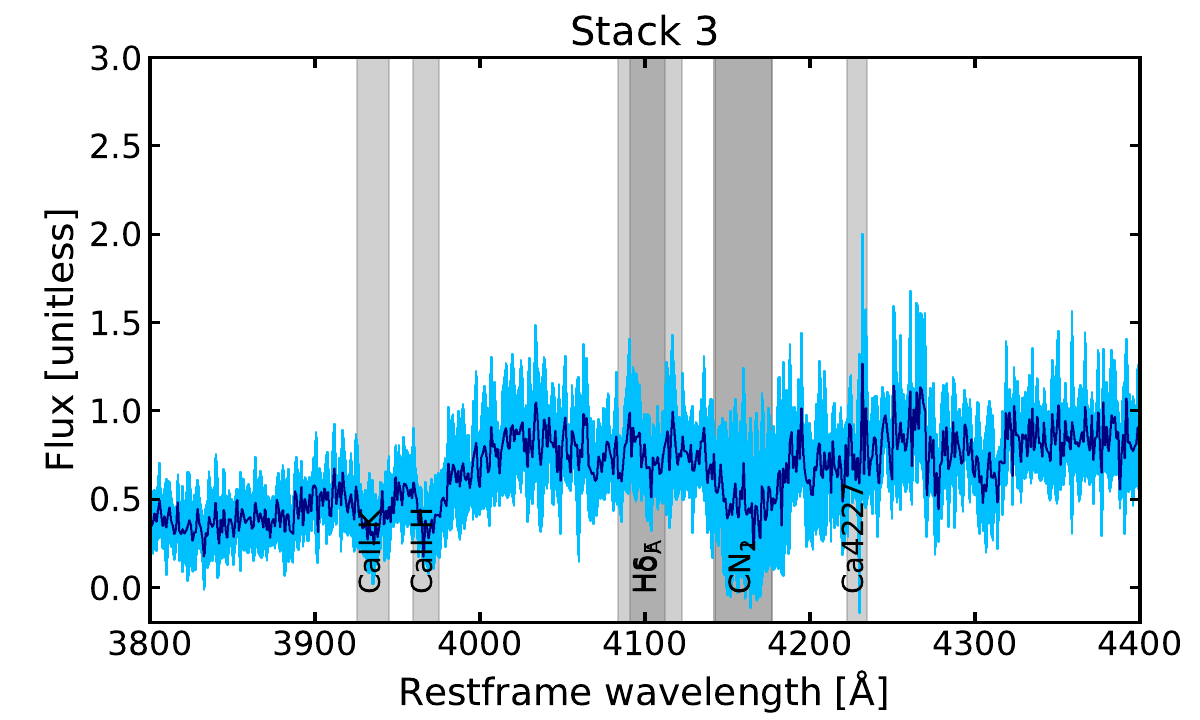}}\\
\subfloat{\includegraphics[scale=0.28]{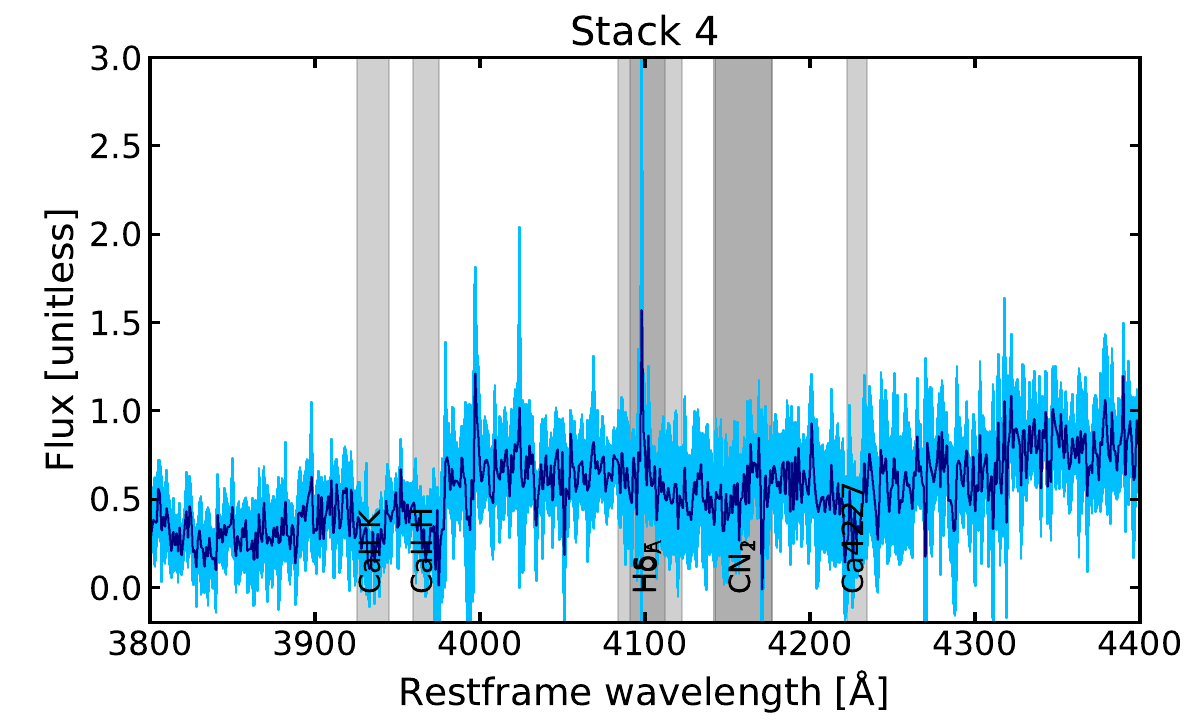}}
\subfloat{\includegraphics[scale=0.28]{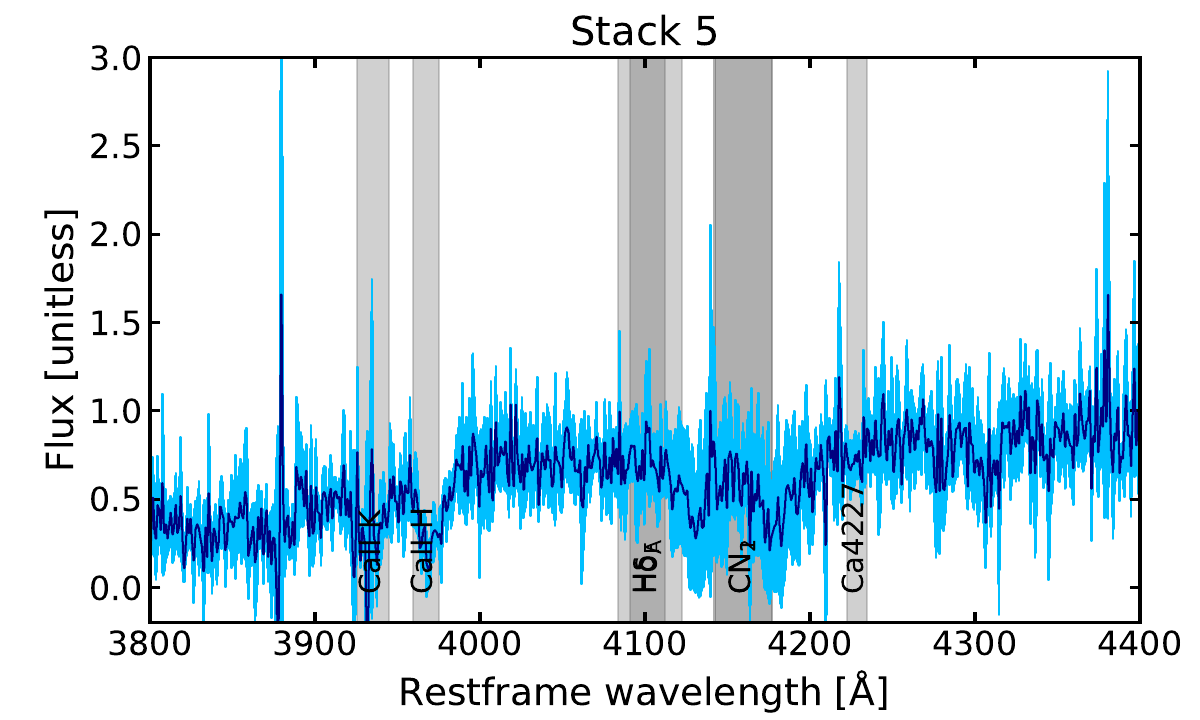}}
\subfloat{\includegraphics[scale=0.28]{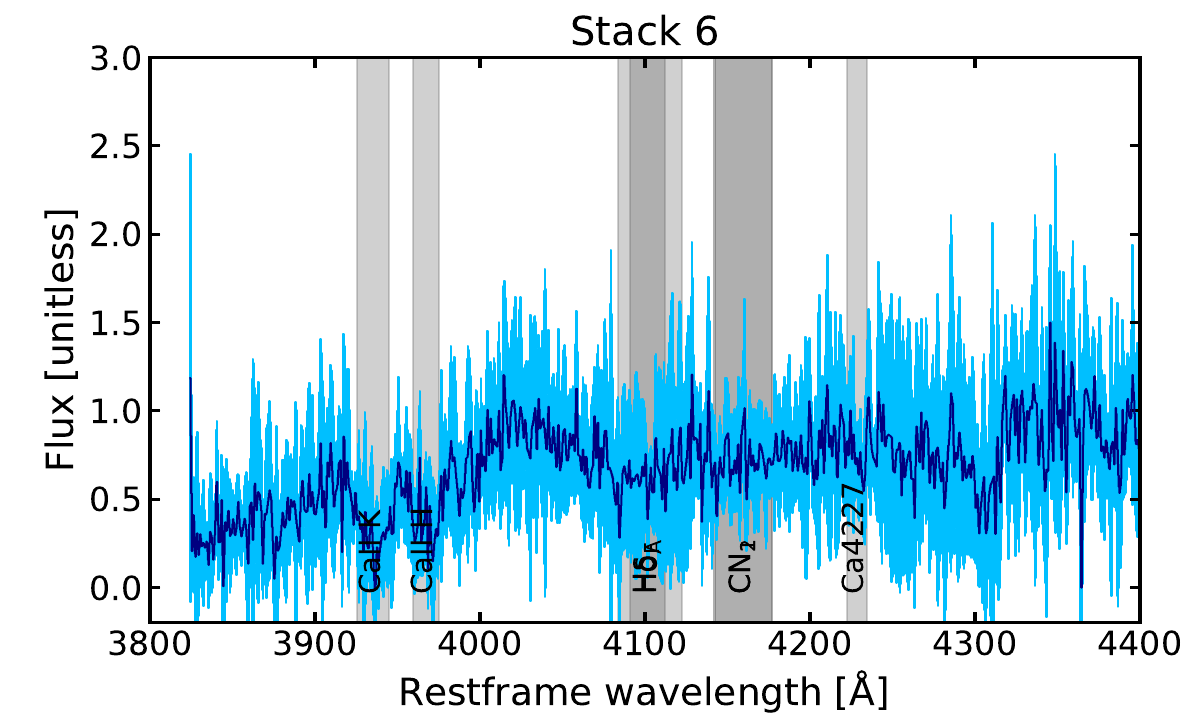}}\\
\subfloat{\includegraphics[scale=0.28]{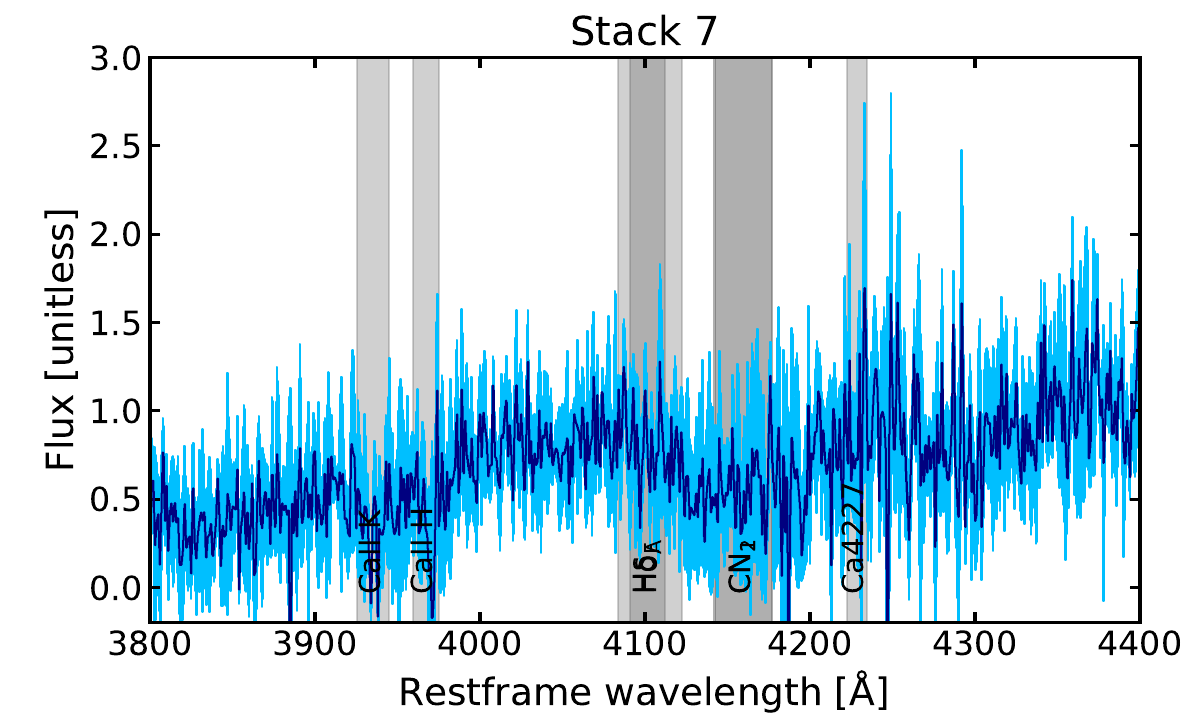}}
\subfloat{\includegraphics[scale=0.28]{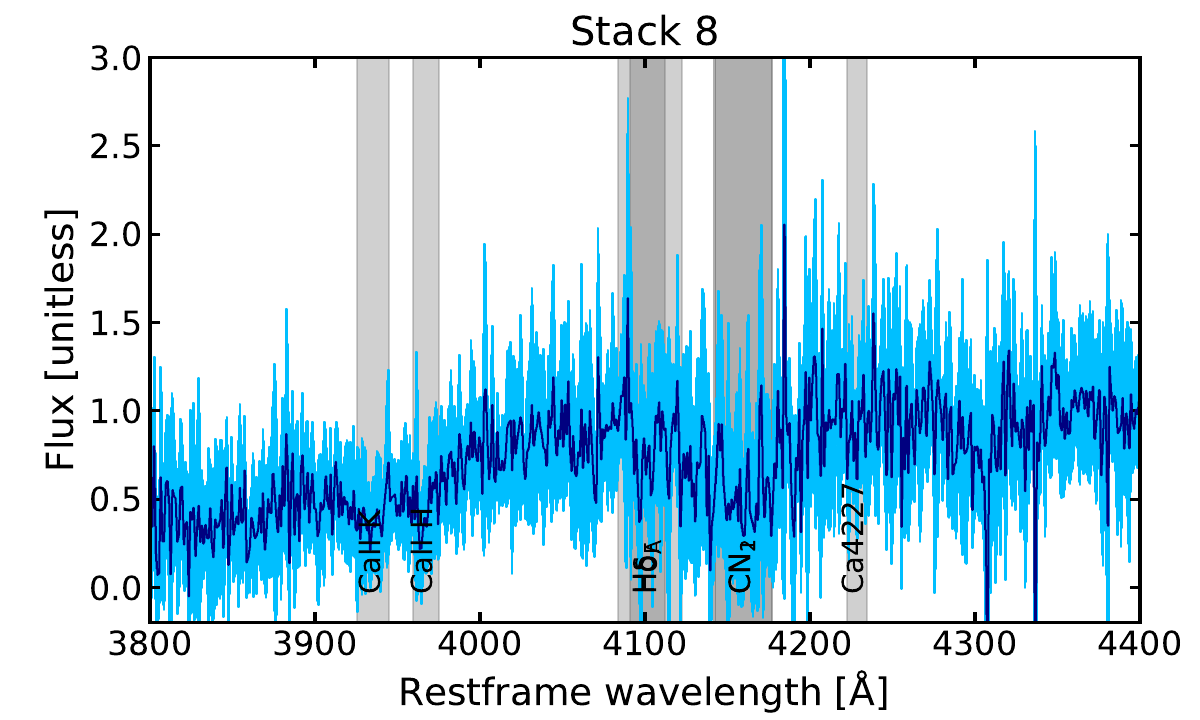}}
\subfloat{\includegraphics[scale=0.28]{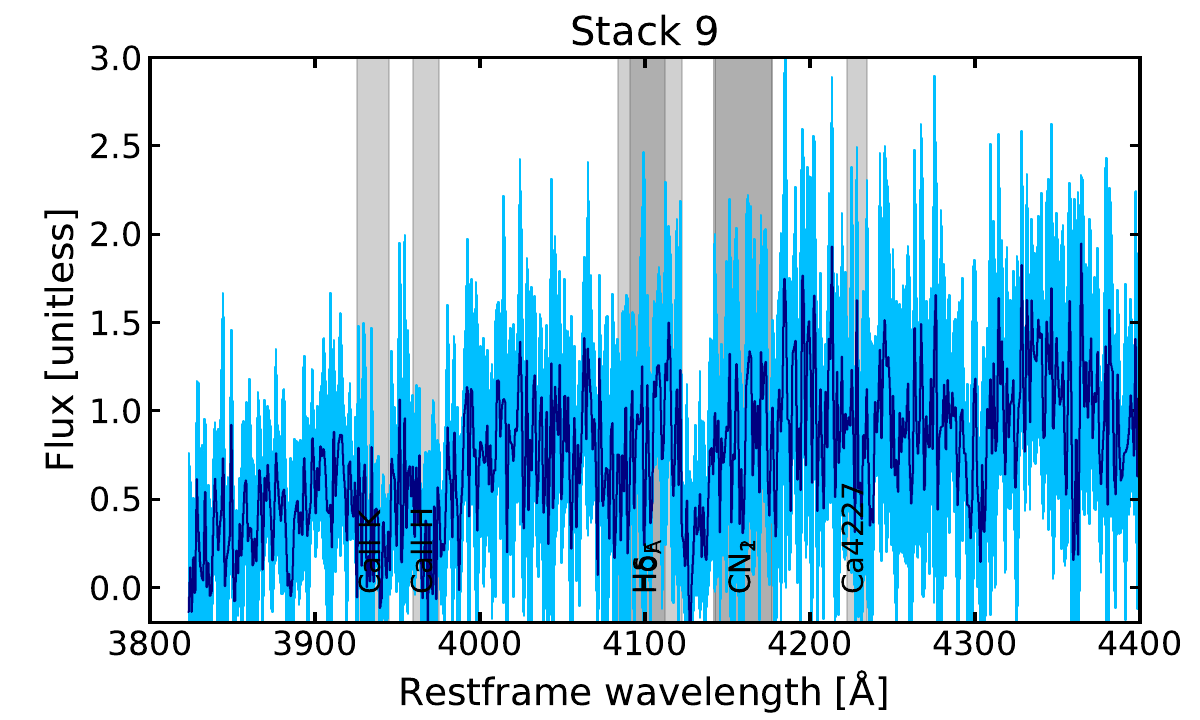}}\\
\subfloat{\includegraphics[scale=0.28]{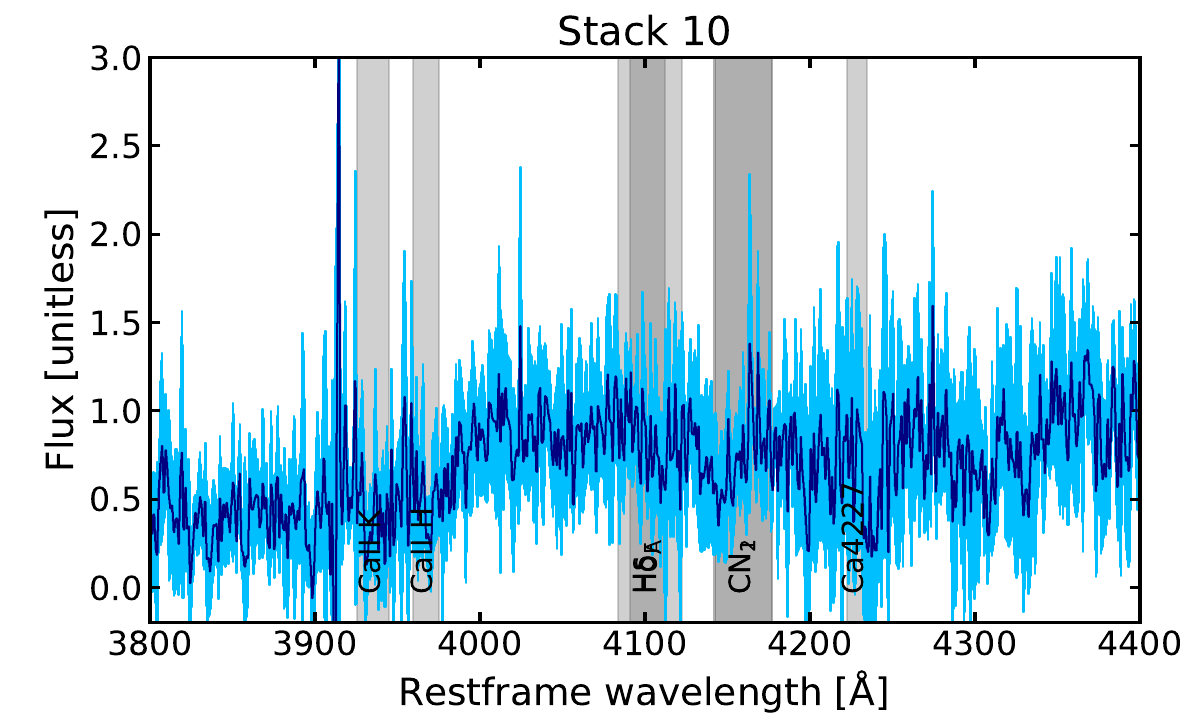}}
\subfloat{\includegraphics[scale=0.28]{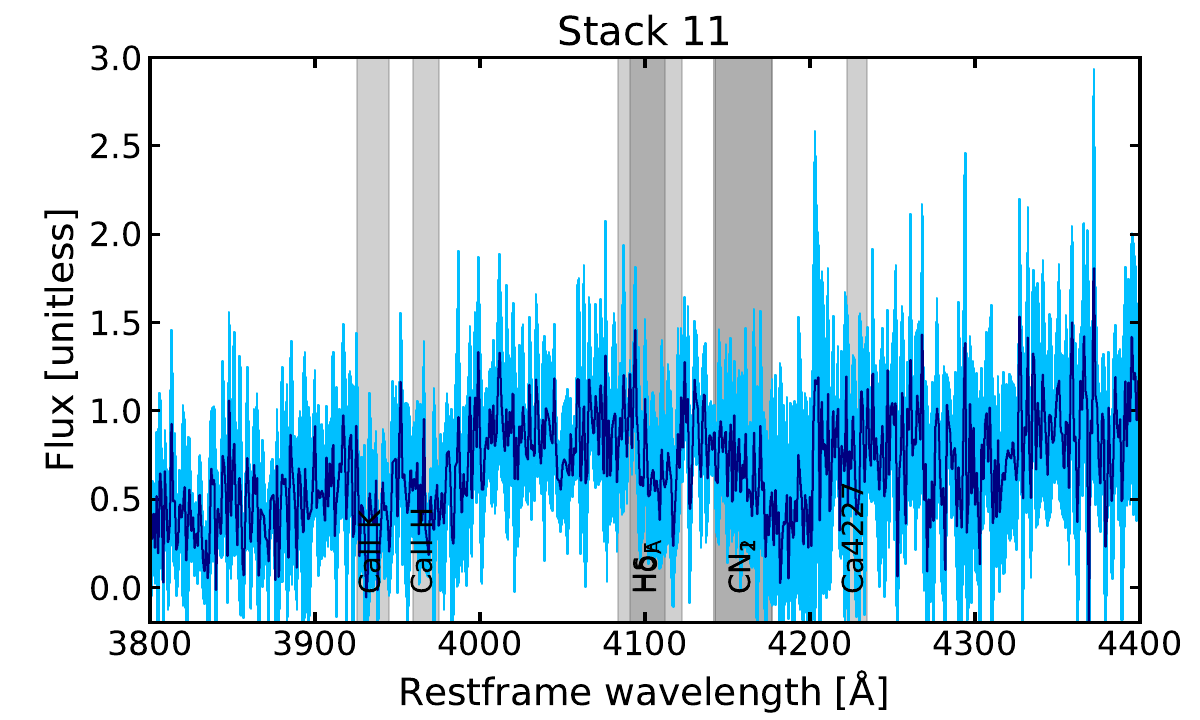}}
\subfloat{\includegraphics[scale=0.28]{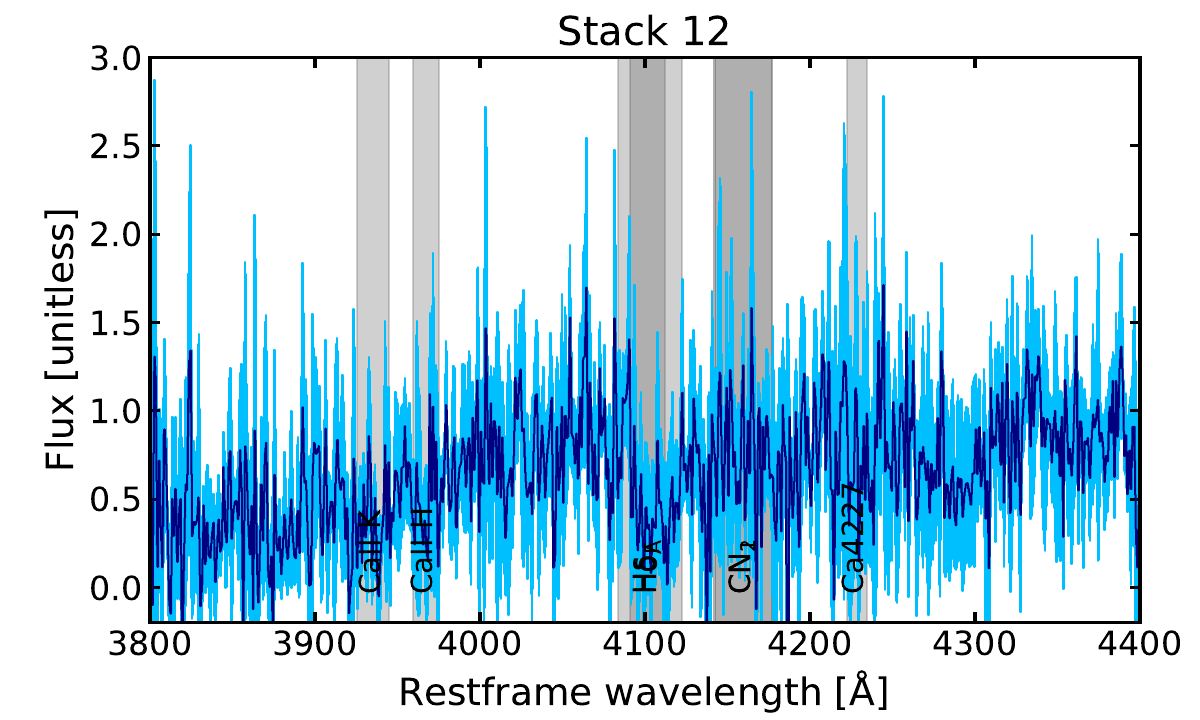}}\\
   \caption{Stacked spectra for the 12 stacks (from the lowest to the highest redshift) given in Table \ref{tableLick}.}
\label{fig:stacked_spectraD4000n}
\end{figure*}

\subsubsection{Ca II K and H}
\label{Cal}

The Ca II K and H lines at 3934 and 3969 \AA, respectively, are two prominent features in the spectra of early-type galaxies. Since the Ca II H line also overlaps with the H$\epsilon$ line of the Balmer series at 3970 \AA{}, the ratio of the two Ca II indices can be used to detect the presence of young star-forming components. \citet{Borghi2022a} and \citet{Jiao2023} define and measure the ratio of Ca II H and K as H/K = I$_{(\rm Ca\ H + H\epsilon)}$/I$_{(\rm Ca\ K)}$. This diagnostic is less sensitive to potential bias introduced by noise peaks in the spectrum with respect to using the H and K flux minima, and has been used to test for the presence of a contaminant young stellar population, therefore, describing the purity of the selected sample \citep{Borghi2022a}. In particular, \citet{Borghi2022a} found that H/K $<$ 1.2 for the spectra of passive galaxies used for CC. Similarly to \citet{Borghi2022b} and \citet{Jiao2023}, we measure the H/K ratios using $\mathtt{PyLick}$. Table \ref{tableLick} shows that all our stacked spectra are H/K $<$ 1.2, with the exception of stack 8 where large errors prevent a concrete conclusion.

\subsubsection{D4000$_{n}$}

\citet{Moresco2011} introduced using the D4000$_{\rm n}$ index for CC. It can be measured at lower S/N and spectral resolution than other spectral indices or full spectrum fitting. This break is a discontinuity of the spectral continuum around 4000 \AA{}, where $n$ indicates that the narrower bands between 3850--3950 \AA{} and 4000--4100 \AA{} are used, which makes the index particularly insensitive to dust \citep{Balogh1999}. We present the measurements of D4000$_{\rm n}$ (using $\mathtt{PyLick}$) in Table \ref{tableLick}. D4000$_{\rm n}$ is also only weakly dependent on the star formation history for old passive stellar populations. 

The break in the continuum is due to metal absorption lines whose amplitude correlates linearly with the metallicity of the stellar populations. The index is also linearly correlated with the age of the stellar population, D4000$_{\rm n}(Z, M)$ = $A(Z, M) \cdot$ age + $B(Z, M)$, so that $dz/dt_{U}$ can be expressed as $A(Z, M) \times (dz/dD4000_{n})$, where the calibration factor $A(Z, M)$, in units of Gyr$^{-1}$, encapsulates dependencies such as metallicity ($Z$) and stellar population modelling ($M$, \citealt{Moresco2011, Moresco2012}). Thus,
\begin{equation}
H(z) = - \frac{A(Z, M)}{1+z} \frac{dz}{dD4000_{n}},
\label{D4000eq}
\end{equation}
where $M$ is a combination of stellar population modelling dependencies such as the IMF, assumed star formation history, stellar library and stellar population models chosen. 

An additional advantage is the decoupling of the statistical uncertainties (all included in the observationally measurable term, $dz/dD4000_{n}$) from systematic effects (captured by the coefficient $A(Z, M)$). Similarly to the method outlined in \citet{Moresco2012}, we will first derive an observed D4000$_{\rm n} - z$ relation and then calibrate the D4000$_{\rm n}$ -- age relation with stellar population synthesis models to quantify the $A(Z, M)$ parameter. We will finally estimate $H(z)$, verifying the robustness of the results against the adopted choice of model ingredients and the stellar metallicity. 

\begin{figure*}
\centering
\includegraphics[scale=0.72]{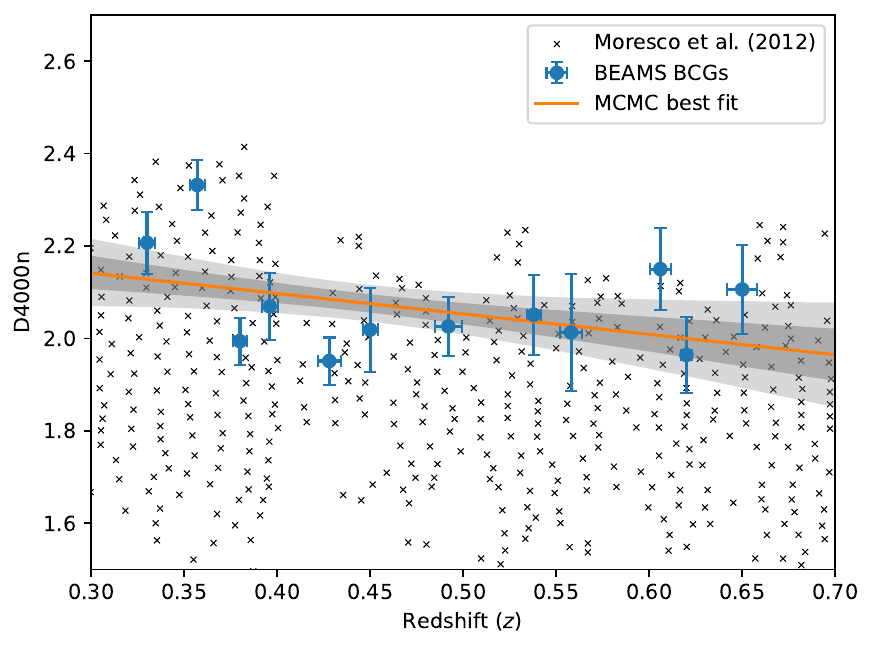}
 \includegraphics[scale=0.7]{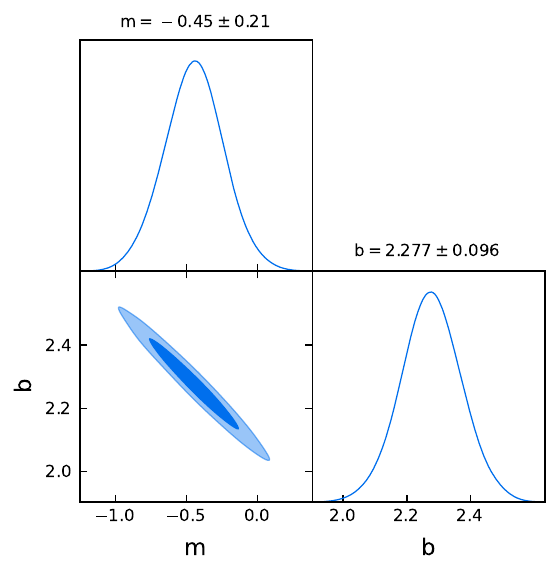}
   \caption{\textit{Left:} The BCG D4000$_{\rm n}$ measurements are shown in blue for the 12 stacked spectra. In the background we also show individual D4000$_{\rm n}$ measurements for early-type galaxies from \citet{Moresco2012} with black crosses. Our BCGs lie at the top end of the data as expected. The line and its surrounding dark gray and shallow gray regions are the best-fitting value of the slope and intercept parameters from the Markov chain Monte Carlo (MCMC), and the associated $\pm 1\sigma$ and $\pm 2\sigma$ confidence range. \textit{Right:} Posterior distribution for slope $m$, and intercept $b$.}
\label{fig:D4000n}
\end{figure*}

\subsubsection{The D4000$_{n} - z$ relation}

We show the BCG D4000$_{\rm n}$ measurements (in blue) against the redshift for the 12 stacked spectra in Figure \ref{fig:D4000n}. For reference we also show the data for a large sample of early-type galaxies from \citet{Moresco2012} with black crosses. The figure shows how the upper envelope of the distribution follows the decrease expected from the ageing of the Universe. Our BCGs lie at the top end of the \citet{Moresco2012} data as expected, as we are probing the upper limits of D4000$_{\rm n}$ \citep{Kauffmann2003, VonderLinden2007, Farage2010, Farage2012}. The figure illustrates the value of using BCGs, the first galaxies in the Universe that end up in the most massive clusters. 

We can assume a linear D4000$_{\rm n} - z$ relation in the relevant redshift range, as long as $H(z)$ can be considered to slowly vary over this redshift range. We derive the mean observed D4000$_{\rm n} - z$ relation for the BCGs that provide the quantity $dz/dD4000_{\rm n}$. We use the median redshift for every bin to plot the data points. It should also be noted that every individual redshift in the stack has an individual uncertainty related to our ability to measure redshift (determined from the resolution of the spectral data). For the resolution of our data, this uncertainty on individual redshifts is between 0.0045 to 0.0055. This is of the same order as the error on the median for the bins (lowest 0.0030 to highest 0.0080). We use the error on the median as the uncertainty on the redshift for each bin. Because the slope of this relation directly determines our $H(z)$ estimate, we use two different methods to estimate its value. 

To account for the uncertainties on both D4000$_{\rm n}$ and redshift, we first use a total least square estimate (TLS, assuming independent Gaussian error bars) and find $m = dD4000_{\rm n}/dz = -0.45 \pm 0.21$ (slope) and $b = 2.277 \pm 0.096$ (intercept). Second, we write down the log-likelihood function as 

\begin{equation}
        \ln \mathcal{L}(m,b) = -\frac{1}{2}\sum^{n=12}_i\left [ \frac{(y_i-mz_i-b)^2}{\sigma^2_i} + {\rm ln} (2\pi \sigma^2_i) \right ], 
        \label{eq:log_likelihood}
\end{equation}
where $y\equiv$D4000$_{\rm n}$ and the summation is over all $12$ D4000$_{\rm n}$ measurements, and the total error represent the error on y (D4000$_{\rm n}$) and redshift. We finally use the Python module \texttt{emcee}\footnote{\url{https://emcee.readthedocs.io/en/stable/}} (5000 steps of the Markov chain Monte Carlo, MCMC), to sample the $m, b$ parameters in Eq.\ \ref{eq:log_likelihood}, and plot the posterior distribution of the two parameters in the right panel of Figure \ref{fig:D4000n}. The best-fitting values are $m = dD4000_{\rm n}/dz = -0.45 \pm 0.21$ and $b = 2.277 \pm 0.096$. This fitting result is almost identical to the TLS estimate, indicating the robustness of the fit. We plot this best-fitting D4000$_{n} - z$ relation with the data in the left panel of Figure \ref{fig:D4000n}. For the D4000$_{\rm n}$ evolution to be larger than the statistical scatter present in the data, we only fit one relation for one measurement at $z=0.5$ for the BCGs. The intercept, $b$, is related to the population of galaxies probed since more massive galaxies are higher within the D4000$_{\rm n}$ -- $z$ envelope (in a fixed redshift range), but is unrelated to our $H(z)$ estimate. 

To summarise, we find the best-fitting slope of the D4000$_{n} - z$ relation as 
\begin{equation}
m = dD4000_{\rm n}/dz = -0.45 \pm 0.21
\label{resultm}
\end{equation}
where the 47\% error is purely statistical. The inverse of Eq.\ \ref{resultm} will be used to calculate the Hubble parameter (Eq.\ \ref{D4000eq}). 

\section{Stellar population models and calibration of the D4000$_{n}$ -- age relation}
\label{calibration}

\begin{figure}
\centering
\includegraphics[scale=0.35]{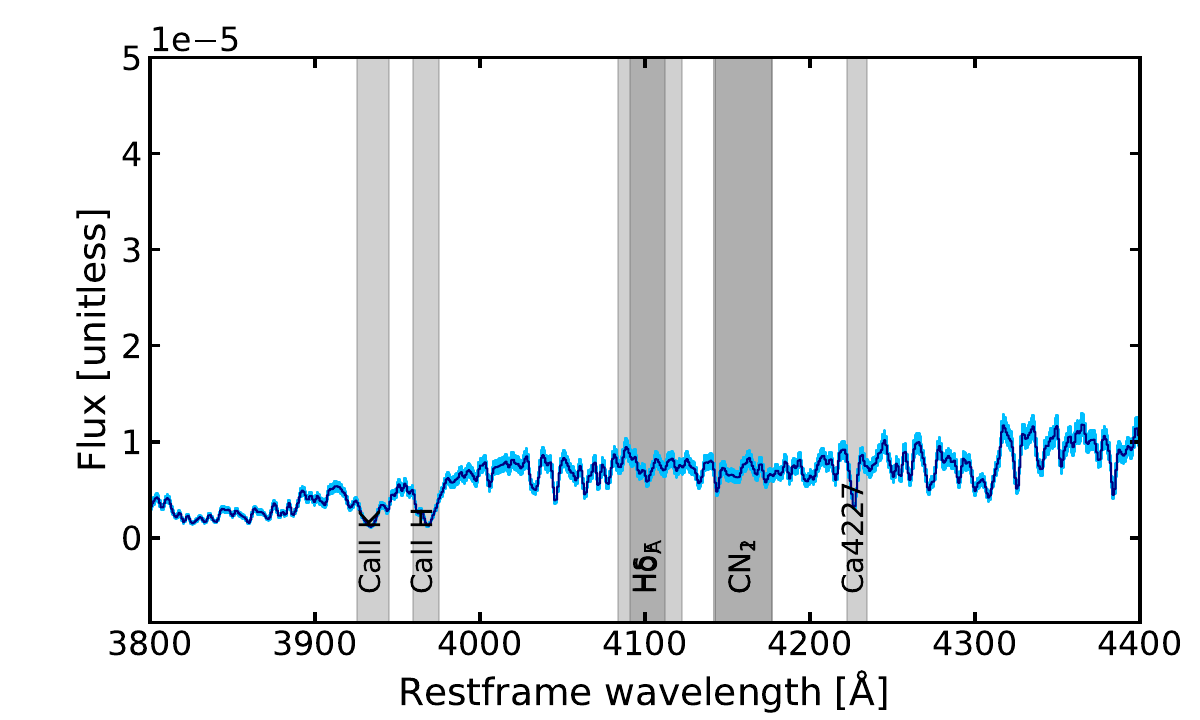}
\caption{We illustrate a mock spectrum generated using the MILES models \citep{Vazdekis2010} with BaSTI isochrones \citep{Hidalgo2018} for a stellar metallicity of $Z$=0.03 and age of 14 Gyr. From the mock spectra we measure D4000$_{\rm n}$ following the identical procedure in $\mathtt{PyLick}$ as for the observational data.}
\label{fig:BaSTI_ex}
\end{figure}

We calibrate the D4000$_{\rm n}$ -- age relation with stellar population synthesis models to quantify the $A(Z, M)$ parameter and its systematic uncertainties. We need to quantify the effect of the stellar metallicity ($Z$), as it has a significant influence on the D4000$_{\rm n}$ -- age relation. We also need to probe the effect of the chosen IMF, the stellar library, and the stellar population models. To do this, we used different stellar population models and ingredients to generate mock spectra, and we measured D4000$_{\rm n}$ following the identical procedure in $\mathtt{PyLick}$ as for the observational data. An example is shown in Figure \ref{fig:BaSTI_ex}. 

\subsection{$A(Z, M)$ for different stellar population models/libraries and different IMFs}
\label{fittingAZ}

\begin{figure}
\centering
\includegraphics[scale=0.29]{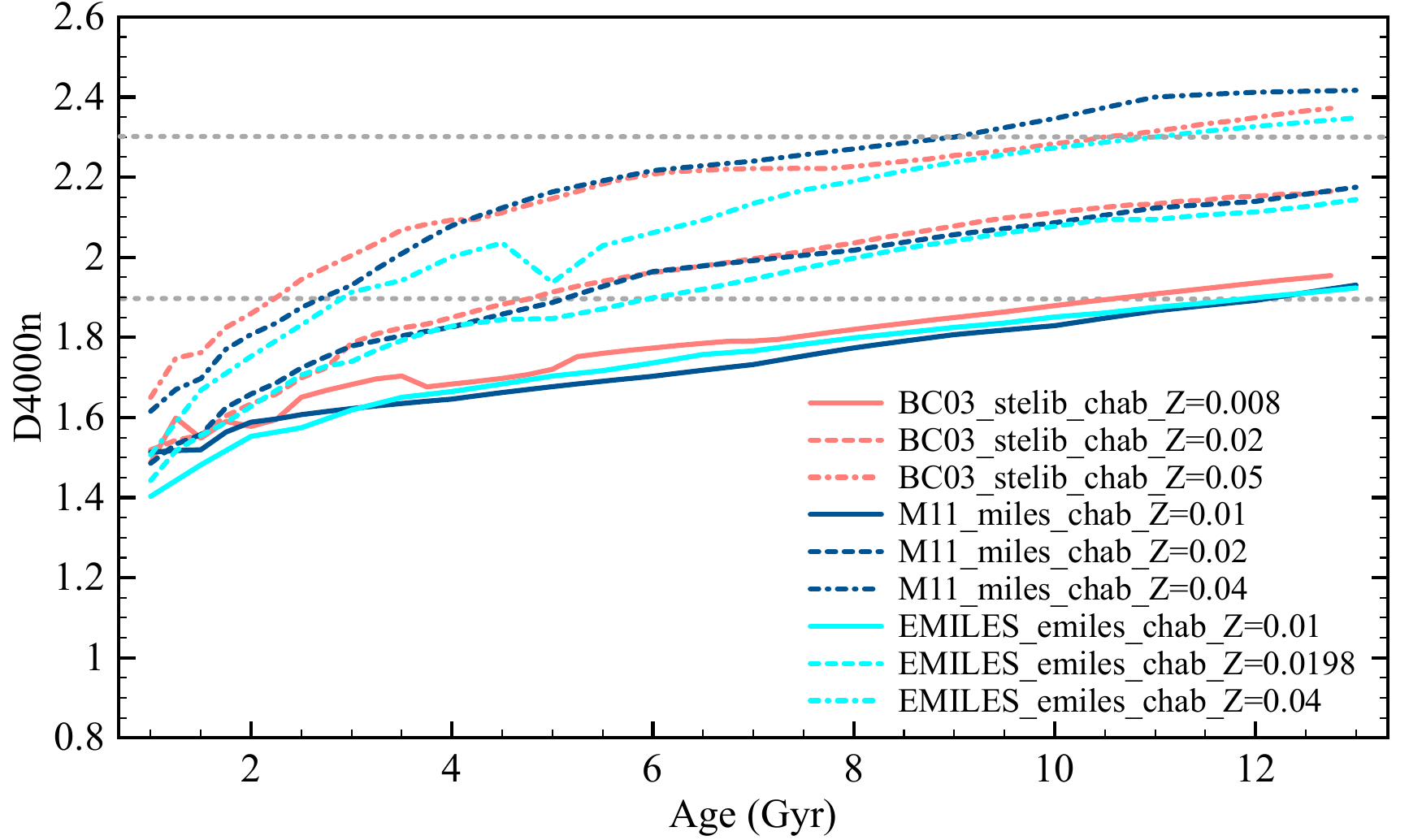}
   \caption{The D4000$_{\rm n}$ -- age relation for three different stellar population models, at three different stellar metallicities, but the same IMF. We show two grey dotted lines to indicate the D4000$_{\rm n}$ measurements that we find for our BCGs (1.9 to 2.3).}
\label{fig:D4000nBC03M11}
\end{figure}

To determine uncertainties that result from the choice of the stellar population synthesis model, we consider different models/libraries. \citet{Moresco2012, Moresco2020} provide a useful, publicly available, stellar population model compilation to derive $A(Z, M)$ parameters. In Figure \ref{fig:D4000nBC03M11}, we show different models at three different stellar metallicities (all with a Chabrier IMF) as an illustration. For the BC03 models \citep{Bruzual2003} we use the STELIB \citep{LeBorgne2004} stellar library with a resolution of 3 \AA{} and a Chabrier IMF with a grid of metallicities Z/Z$_{\sun}$ = [0.4, 1.0, 2.5]. The MaStro (M11) models \citep{Maraston2011}, with a resolution of 2.3 \AA{}, have been constructed using the MILES \citep{Vazdekis2010} stellar library and a Chabrier IMF and three stellar metallicities, namely Z/Z$_{\sun}$ = [0.5, 1.0, 2.0]. For the E-MILES models, we use BaSTI isochrones \citep{Hidalgo2018}, the MILES stellar library, a Chabrier IMF, and the stellar metallicities Z/Z$_{\sun}$ = [0.5, 0.99, 2.0]. The strong dependence of $A(Z, M)$ on stellar metallicity can be seen in Figure \ref{fig:D4000nBC03M11}. We also show two grey dotted lines to indicate the range of D4000$_{\rm n}$ measurements that we find for our BCGs (1.9 to 2.3). The systematic errors using different models are further quantified in Section \ref{systematics}.

\begin{figure}
\centering
\includegraphics[scale=0.29]{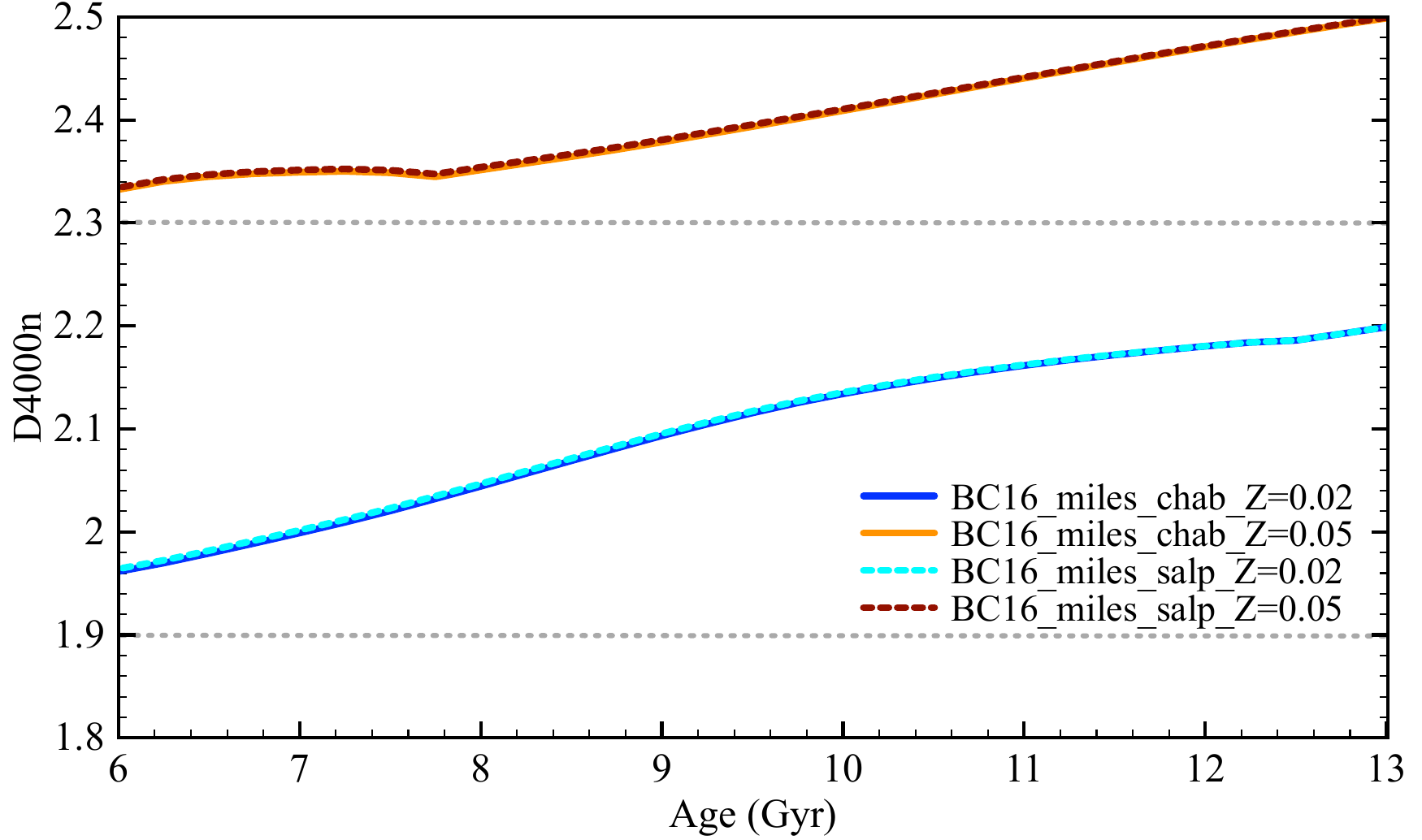}
   \caption{The BC16/MILES models for Chabrier and Salpeter IMFs at $Z$ = 0.02 and 0.05, to illustrate the negligible effect of the IMF.}
\label{fig:D4000nBC16IMF}
\end{figure}

We also consider the sensitivity of $A(Z, M)$ to the choice of the IMF. In Figure \ref{fig:D4000nBC16IMF}, as an illustration, we show the updated BC16 models \citep{Charlot2016} with the MILES library \citep{Vazdekis2010} with a resolution of 2.3 \AA{}, at two stellar metallicities ($Z$ = 0.02 and 0.05), for both a Chabrier and a Salpeter IMF. We confirm that different IMFs have negligible influence on $A(Z, M)$.

\subsection{Quantifying the dependence of $A(Z, M)$ on stellar metallicities for BCGs}
\label{metdependence}

\begin{figure}
\centering
\includegraphics[scale=0.30]{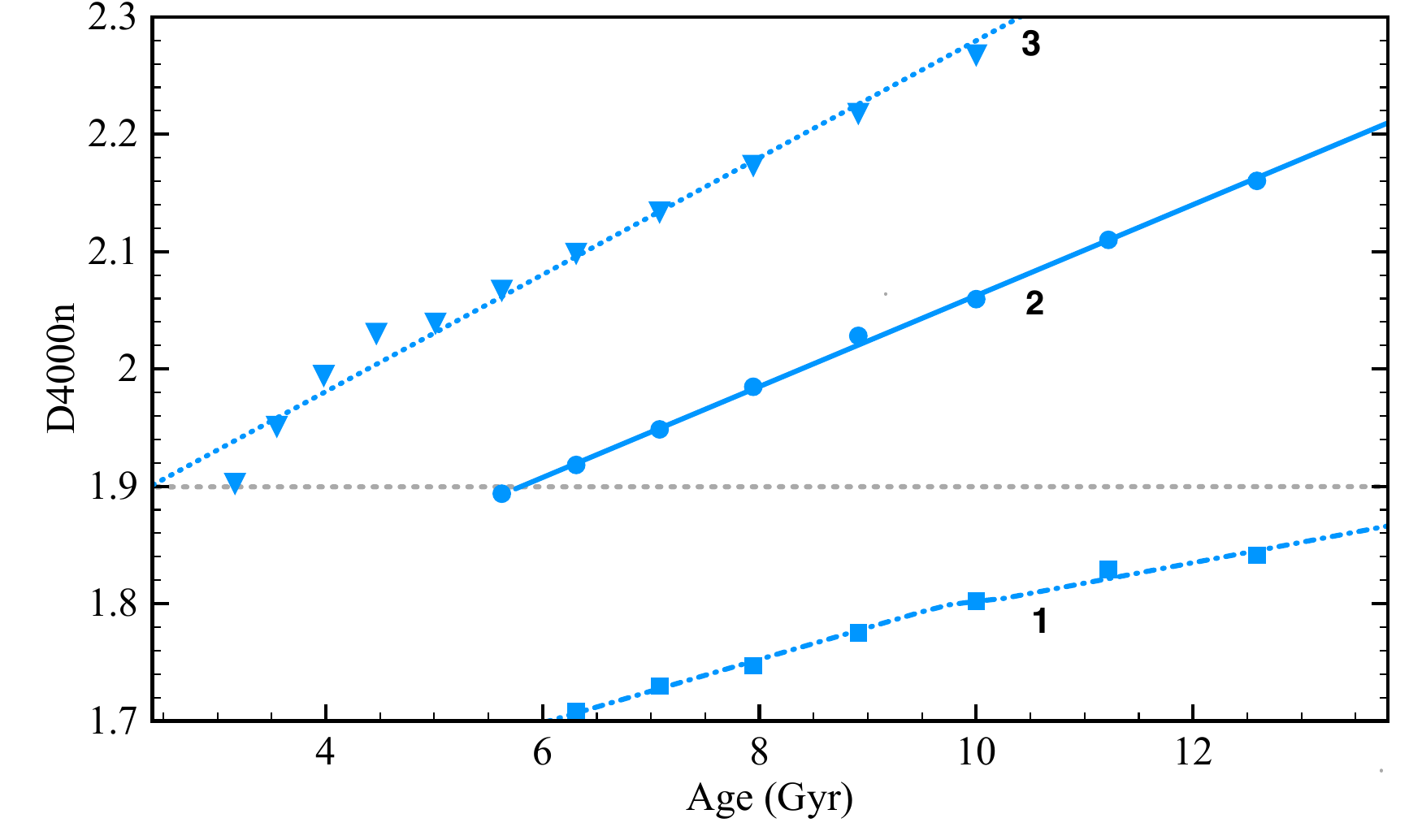}
   \caption{The D4000$_{\rm n}$ -- age relation for MILES models with Padova isochrones and a Chabrier IMF for metallicities $Z$ = 0.008, $Z$ = 0.019 (solar metallicity) and $Z$ = 0.030 (indicated by 1, 2, and 3, respectively) fitted with lines as described in Section \ref{metdependence}.}
\label{fig:D4000nMILESzoom}
\end{figure}

After investigating all possible models, we choose the MILES models with Padova isochrones \citep{Girardi2000} to quantify $A(Z, M)$ as a function of stellar metallicities for BCGs, as the D4000$_{\rm n}$ -- age relation for these models can be approximated with a single straight line between D4000$_{\rm n}$ from 1.9 to 2.3. We used the public available webtool for MILES\footnote{\url{http://research.iac.es/proyecto/miles/pages/webtools/tune-ssp-models.php}} to obtain models with Padova isochrones, a Chabrier IMF, and spectral resolution that matches that of our observed spectra, and measure D4000$_{\rm n}$ with $\mathtt{PyLick}$. In Figure \ref{fig:D4000nMILESzoom}, we show the relation for stellar metallicities at $Z$ = 0.008, $Z$ = 0.019 (solar metallicity) and $Z$ = 0.030. For the relations at $Z$ = 0.019 and $Z$ = 0.030, the D4000$_{\rm n}$ measurements are in our range of interest (from 1.9 to 2.3), and we approximate the relations with straight lines as shown in Figure \ref{fig:D4000nMILESzoom} (indicated by the numbers 2 and 3). We show the derived values of $A(Z, M)$ in Table \ref{tableAZ}, together with the corresponding values of $H(z)$. 

\begin{table*}
\caption{To quantify the systematic error due to stellar metallicity, we give the slope $A(Z, M)$ and the corresponding $H(z)$ for each stellar metallicity. For the $H(z)$ calculation we use $dD4000_{\rm n}/dz$ as --0.45 $\pm$ 0.21. The three line segments are marked in Figure \ref{fig:D4000nMILESzoom}.}    
\label{tableAZ}      
\centering                         
\begin{tabular}{l c c c} 
\hline
Model & Regime & $A(Z, M)$ & $H(z)$ in km s$^{-1}$ Mpc$^{-1}$ \\
\hline                
MILES, Padova, chabrier IMF, Z=0.5 Z$_{\sun}$ (Z=0.008) & 1) $1.8 < D4000_{\rm n} <$ 1.9 &	0.0174	&  25.2 $\pm$ 11.8 \\
MILES, Padova, chabrier IMF, Z=1.0 Z$_{\sun}$ (Z=0.019) & 2) $1.9 < D4000_{\rm n} <$ 2.3       & 0.0388	 &  56.2 $\pm$ 26.4 \\
MILES, Padova, chabrier IMF, Z=1.5 Z$_{\sun}$ (Z=0.030) & 3) $1.9 < D4000_{\rm n} <$ 2.3   &	0.0498	 &  72.1 $\pm$ 33.9 \\
\hline
\end{tabular}
\end{table*}

For $Z$ = 0.008, the D4000$_{\rm n}$ values are not in our range (above 1.9), so we fit the relation with a piecewise linear slope adopting the public Python code $\mathtt{pwlf}$\footnote{\url{https://pypi.org/project/pwlf/}} \citep{pwlf}, and we use the segment of the line just below 1.9 (D4000$_{\rm n}$ from 1.8 to 1.9, indicated by number 1 in Figure \ref{fig:D4000nMILESzoom}) to get a lower reference point in Table \ref{tableAZ}. This choice does not influence the final result for $H(z)$ in this study.    

At low redshifts ($z < 0.03$), BCGs have metallicities that are on average twice solar \citep{Loubser2009}, as the metallicity also increases with mass. Comparable results were obtained by \citet{Groenewald2014}, and for samples at higher redshifts by \citet{Lidman2012, Bellstedt2016}, and for simulations by \citet{Contreras-Santos2022}. BCGs are more metal-rich than other early-type galaxies which can have $Z/Z_{\sun} \sim 1 - 1.5$ (see \citealt{Moresco2012, Moresco2016}). On a theoretical basis, for galaxies that have fully exhausted their gas reservoir and completed their mass assembly at high redshifts and have evolved passively since then, a negligible evolution in stellar metallicity is expected. 

\begin{figure}
\centering
\includegraphics[scale=0.30]{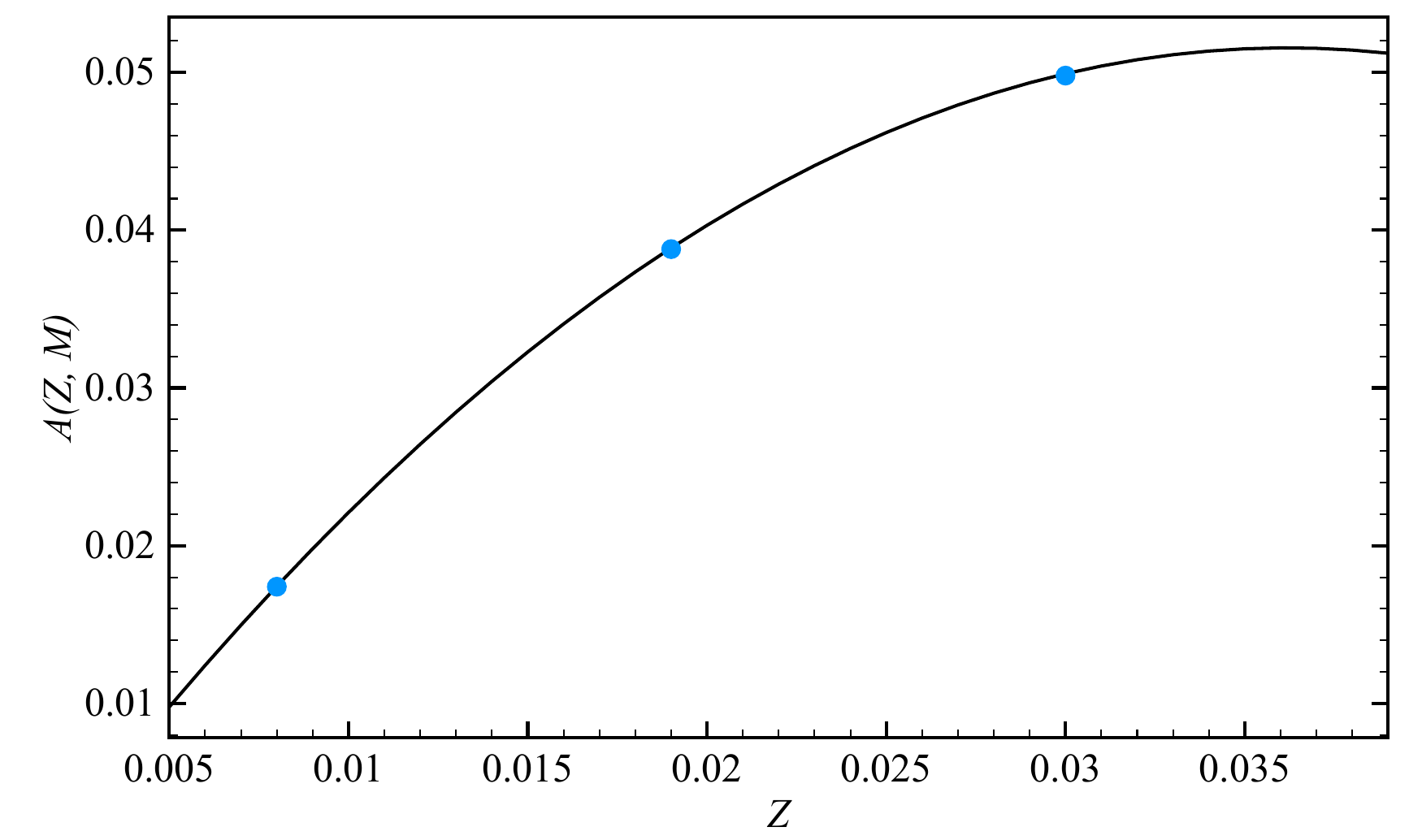}
   \caption{$A(Z, M)$ as a function of $Z$ using line segments 1 ($1.8 < D4000_{\rm n} <$ 1.9, $Z$ = 0.008), 2 ($1.9 < D4000_{\rm n} <$ 2.3, $Z$ = 0.019) and 3 ($1.9 < D4000_{\rm n} <$ 2.3, $Z$ = 0.030) in Figure \ref{fig:D4000nMILESzoom}. $A(Z, M)$ is much less sensitive to $Z$ at the typical metallicities of BCGs (compared to solar metallicities).}
\label{fig:AZvsZ}
\end{figure}	 

In Figure \ref{fig:AZvsZ} we investigate the sensitivity of $A(Z, M)$ with $Z$. We show the $A(Z, M)$ calibrations at $Z$ = 1.0 $Z_{\sun}$ and $Z$ = 1.5 $Z_{\sun}$ for $1.9 < D4000_{\rm n} <$ 2.3. We also have a reference point at lower $Z$ (0.5 $Z_{\sun}$) where we have to use $1.8 < D4000_{\rm n} <$ 1.9. We can fit a second-order polynomial and use it to extrapolate to $Z$ = 0.038 (2.0 Z$_{\sun}$), but the $A(Z, M)$ values converge at higher metallicity as shown in Figure \ref{fig:AZvsZ}. $A(Z, M)$ is much less sensitive to $Z$ at the typical metallicities of BCGs (compared to solar metallicities) and is unlikely to be higher than 0.0514. 

\section{Systematic errors and comparisons}
\label{systematics}

We verify the robustness of the results of $H(z)$ against the adopted choice of stellar population synthesis model, star formation history, IMF and the sensitivity to uncertainties in stellar metallicity. We also compare our results with those of earlier studies. 

\subsection{Systematic errors}

Some previous studies \citep{Moresco2020, Borghi2022b} focus on an extensive analysis of the different contributions to the systematic error budget ($\sigma_{\rm syst}$). Through simulations ($z < 1.5$), \citet{Moresco2020} find that the choice of the stellar population synthesis model contributes 5\% to the total error budget on $H(z)$. The contribution due to the choice of the IMF is smaller than 1\%, while that due to the stellar library is 7\%, on average. \citet{Moresco2020} also assessed the impact of an uncertainty on the stellar metallicity determination, finding that an error of 10\% on the stellar metallicity propagates to a 9\% error on $H(z)$. We briefly discuss how the different factors affect the BCG measurements in particular. 

\subsubsection*{Star formation history (SFH) and contamination of spectra by young stellar components}

Late star formation in quiescent massive galaxies can lead to a systematic overestimation of the Hubble parameter \citep{Liu2016}. We expect BCGs to have a SFH with a very small duration ($< 0.3$ Gyr) that can be parameterised with an exponentially declining SFH, and stellar populations very close to an SSP \citep{Thomas2010, McDermid2015}. \citet{Moresco2012, Moresco2016} demonstrated that the uncertainty on the SFH of passive early-type galaxies impacts the estimate of $H(z)$ at a 2 -- 3\% level at most. 

Our stacked spectra satisfy the criteria determined by \citet{Borghi2022a} of the Ca II line ratio H/K < 1.2 (to avoid a < 200 Myr stellar component). We also did not detect any emission lines during our $\mathtt{pPXF}$ analysis in Section \ref{fitting}. In particular, the $[$O II$]$ and H$\alpha$ emission lines are extremely sensitive to the presence of the youngest and hottest stars (see also \citealt{Loubser2024}). In \citet{Loubser2016}, we analysed long-slit spectra for 19 BCGs (0.15 $< z <$ 0.30) in X-ray luminous clusters. We fitted young stellar components superimposed on an intermediate/old stellar component to accurately constrain their star formation histories. We detected young ($\sim$ 200 Myr) components in four of the 19 BCGs. These four BCGs also showed clear emission lines, blue colours in their photometry, and most importantly, all showed D4000$_{\rm n} < 1.6$ (their figure 1). It is very unlikely that BCGs with D4000$_{\rm n} > 1.9$ (at $z > 0.3$) contain any significant young stellar population component. For our analysis, spectra contamination by a younger stellar component has a negligible influence on the systematic error.  

\subsubsection*{IMF}
Figure \ref{fig:D4000nBC16IMF} shows that the impact of the IMF is negligible. This also confirms previous findings, for example \citet{Moresco2012} who found that the impact of the IMF on the D4000$_{\rm n}$ -- age relation is insignificant in that the difference between D4000$_{\rm n}$ values estimated in a stellar population model with a Chabrier or a Salpeter IMF is less than 0.3\% at all metallicities.

\subsubsection*{Stellar population model and stellar library}
The extensive analysis of the uncertainty of the stellar population model by \citet{Moresco2020}, who used many different models over a large redshift range, quantified that the bias in $H(z)$ due to the stellar library is 6 to 7\% in our redshift range of interest. They also found that the bias due to stellar population models is 10 to 13\% when all the models are considered, but decreases to 5 to 6\% when the biggest outlier is excluded. The error resulting from the different stellar population models adopted is in many cases mainly driven by a single model significantly different from the others.

\subsubsection*{Stellar metallicity (Z)}

We measure the stellar metallicity of our stacked spectra using full-spectrum fitting in Section \ref{fitting}. The results can be seen in Figure \ref{fig:stacked_spectra_ppxf}, and even though the full-spectrum fitting results should be interpreted with care, it suggests that the metallicity of all the stacked spectra is at the upper bound of the models, i.e., $[$M/H$] = 0.4$ (i.e., 2.5 times that of the Sun). However, there is some uncertainty about the results of the full-spectrum fitting, and therefore we also look at the sensitivity of $A(Z, M)$ to $Z$ in Section \ref{metdependence}. At high values for D4000$_{\rm n}$, the dependence on $Z$ converges to a shallow slope. This makes $H(z)$ obtained from BCGs far less sensitive to uncertainties in $Z$ than for less massive luminous red galaxies. Using the MILES models, Padova isochrones, and Chabrier IMF, we find $H(z)$ = 56.2 km s$^{-1}$ Mpc$^{-1}$ at 1.0 $Z_{\sun}$ and $H(z)$ =  72.1 km s$^{-1}$ Mpc$^{-1}$  at 1.5 $Z_{\sun}$, with a smaller difference at 2.0 $Z_{\sun}$ where according to Figure \ref{fig:AZvsZ}, $H(z)$ can be estimated at a maximum of 74.5 km s$^{-1}$ Mpc$^{-1}$ if $A(Z, M)$ is 0.0514. We use the value of $H(z)$ =  72.1 km s$^{-1}$ Mpc$^{-1}$ at 1.5 $Z_{\sun}$ for further calculations, and a maximum systematic error of 3\% due to metallicity. 

\subsection{Total error}

For systematic errors, the conservative total systematic error $\sigma_{\rm syst}$ is:
\begin{equation}
\sigma_{\rm syst} = \pm 3 \% (\rm SFH) \pm 7 \% (\rm library) \pm 6 \% (\rm model) \pm 3 \% (Z),
\end{equation} 
thus 10.1\% when added in quadrature. Compared to Eq. \ref{resultm}, the systematic error is only a quarter of the statistical error, therefore, our measurement of $H(z)$ is completely dominated by the statistical error and the final result is not very sensitive to the calibration of $A(Z, M)$. By combining with Eq.\ \ref{resultm}, we quote our final numerical result as
\begin{equation}
H(z=0.5) = 72.1 \pm 33.9(\rm stat) \pm 7.3 (\rm syst)\ \rm km\ \rm s^{-1}\ \rm Mpc^{-1}.
\end{equation} 

\subsection{Comparison with previous measurements of Cosmic Chronometers (CC)}
\label{previousCC}

\begin{figure*}
\centering
\includegraphics[scale=1.0]{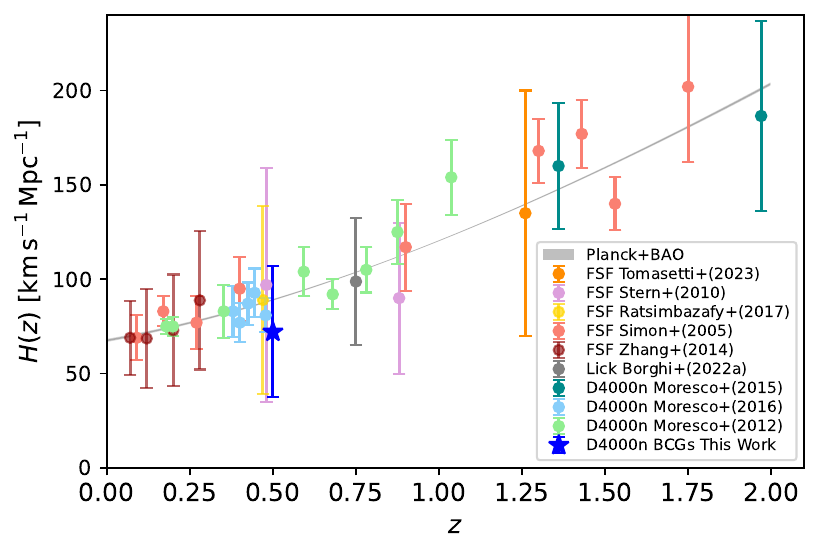}
   \caption{$H(z)$ derived from the cosmic chronometers method at different redshifts. We use the data compiled by \citet{Moresco2022}, and add the data from \citet{Tomasetti2023}. Our estimate using only BCGs is shown with a blue star at $z=0.5$. We indicate whether the measurements were made using full-spectrum fitting (FSF), Lick indices, or D4000$_{\rm n}$. For reference, we show the standard $\Lambda$CDM $Planck$+BAO $H(z)$ evolution in a gray band with its width determined by $\pm 1 \sigma$ variation of cosmological parameters (\citealt{Alam2021, Tang2024}, see also Section \ref{projection}).}
\label{fig:Hz}
\end{figure*}

The CC method has been used successfully up to $z \sim 2$. We summarise previous results and present them, colour-coded by method, together with our estimate in Figure \ref{fig:Hz}. The previous results were obtained using samples of early-type galaxies, not specifically only BCGs. 

\citet{Moresco2012} present eight constraints on $H(z)$ over a wide redshift range of 0.15 $< z <$ 1.1, obtained from a large sample of early-type galaxies ($\sim$ 11000) from several spectroscopic surveys. Later, \citet{Moresco2016} exploited the statistics provided by the Baryon Oscillation Spectroscopic Survey (BOSS) Data Release 9 to extract a sample of more than 130000 massive and passively evolving galaxies, obtaining new $H(z)$ measurements in the redshift range 0.3 $< z <$ 0.5, in particular $H(z = 0.43) = 91.8 \pm 5.3$ km s$^{-1}$ Mpc$^{-1}$. \citet{Borghi2022b} analyze the stellar ages obtained from a combination of Lick indices for 140 massive and passive galaxies selected in the LEGA-C survey at 0.6 $< z < $ 0.9, to obtain $H(z = 0.75)$ = 98.8 $\pm$ 33.6 km s$^{-1}$ Mpc$^{-1}$. At the same redshift, \citet{Jimenez2023} presents a determination of $H(z)$ from photometric data, obtaining $H(z = 0.75) = 105.0 \pm 7.9 (\rm stat) \pm 7.3 (\rm syst)$ km s$^{-1}$ Mpc$^{-1}$. \citet{Jiao2023} perform full-spectrum fitting of 350 massive and passive galaxies selected from the LEGA-C ESO public survey to derive $H (z = 0.8) = 113.1 \pm 15.1(\rm stat) ^{+29.1} _{-11.3}(\rm syst)$ km s$^{-1}$ Mpc$^{-1}$. At even higher redshifts, \citet{Tomasetti2023} obtain $H(z = 1.26) = 135 \pm 65 (\rm stat\ and\ syst)$ km s$^{-1}$ Mpc$^{-1}$. We present a full comparison of the CC results, colour-coded by analysis method, in Figure \ref{fig:Hz}. 

\subsection{$H_{0}$ projection}
\label{projection}

We now project our measurement of $H(z=0.5)$ to $z=0$, and obtain the value of $H_0$. In the standard flat $\Lambda$CDM model, 
\begin{equation}
H(z) = H_{0} \sqrt{\Omega_{\rm m} (1 + z)^{3} + \Omega_{\Lambda}}, \label{eq:Hz-eq}
\end{equation}
where $\Omega_{\rm m}$ is the fractional matter density and $\Omega_{\Lambda}$ is the fractional density of the cosmological constant. We use the {\it Planck}+BAO16 posterior distributions of $\Omega_{\rm m}$ and $\Omega_{\Lambda}$ as priors to divide the redshift-dependent term on the right-hand side of Eq.\ \ref{eq:Hz-eq} by the left to obtain $H_{0}$. The {\it Planck}+BAO16 is the combined dataset of the CMB high-$\ell$ temperature (TT), high-multipole polarisation (TE and EE) {\tt plik} likelihood, low-$\ell$ temperature {\tt Commander} likelihood, and low-$\ell$ {\tt SimAll} EE polarization from the {\it Planck} 2018 data release~\citep{plc18-lk}, and the {\it Planck} 2018 CMB lensing power spectrum~\citep{plc18-ls}, as well as the BAO measurement from SDSS-IV DR16 eBOSS Luminous Red Galaxies (LRG, \citealt{Alam2021}). The distributions are
\begin{eqnarray}
    \Omega_{\rm m} &=& 0.3122 \pm 0.0057 \nonumber  \\
    \Omega_{\Lambda} &=& 0.6878 \pm 0.0057. 
    \label{eq:prior-m-lambda}
\end{eqnarray}
Then the result of the projection is
\begin{equation}
H_{0}=54.6\pm 25.7\,({\rm stat})\pm 5.5 ({\rm syst})\ \rm km\ s^{-1}\ Mpc^{-1}.
\end{equation}

The extrapolation of $H_{0}$ inherently depends on the assumed cosmological model; however, this dependence is a well-established aspect of cosmology and remains relatively stable given the stringent constraints on the matter density parameter, $\Omega_{\rm m}$. In equation \ref{eq:Hz-eq}, $\Omega_{\rm m}$ is tightly constrained to 0.3122 $\pm$ 0.0057, limiting its possible variation. More recent results from the ACT DR6 have further refined this constraint to $\Omega_{\rm m}$ = 0.3032 $\pm$ 0.0048 \citep{Louis2025}.

Importantly, these constraints are not derived solely from {\it Planck} satellite data but also from independent observations, including those from ACT and the South Pole Telescope (SPT). Given this convergence of evidence from multiple independent surveys, the prior on $\Omega_{\rm m}$ can be considered robust. Due to the small uncertainty in $\Omega_{\rm m}$, its contribution to the overall error propagation in $H_{0}$ remains minimal, with the dominant source of uncertainty arising from the direct measurement of $H(z)$.

\subsubsection*{Early-Universe measurements}

\begin{figure*}
\centering
\includegraphics[scale=1.2, trim={0 40 0 30}, clip]{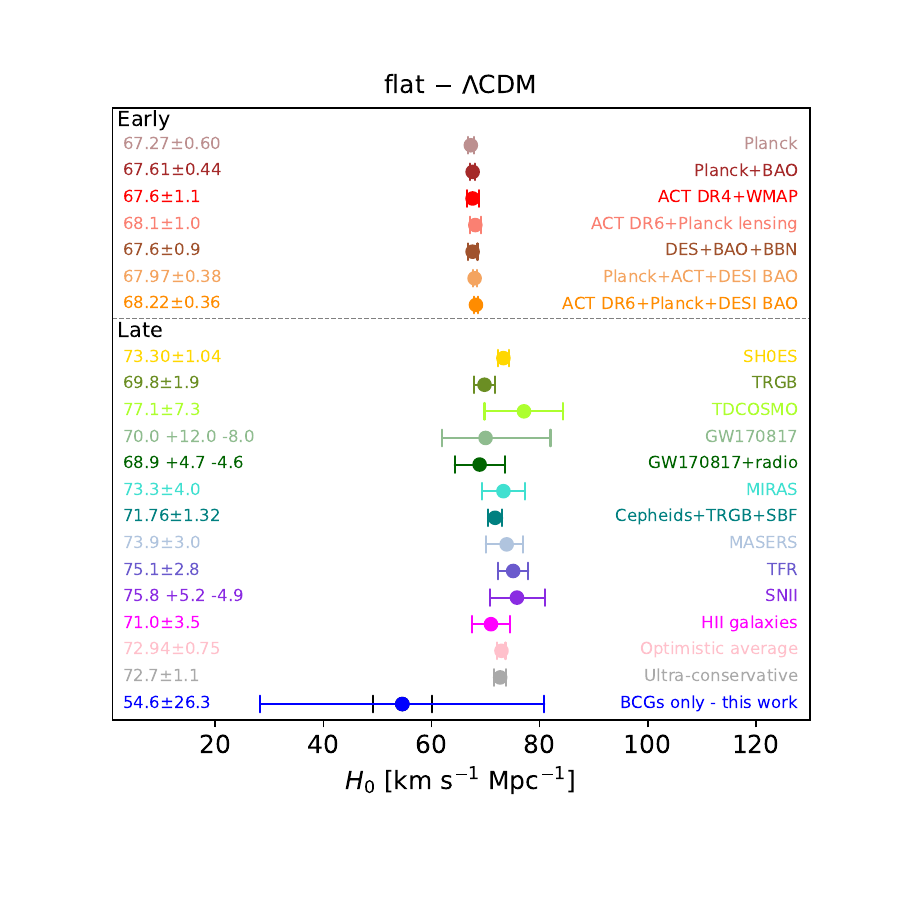}
   \caption{$H_{0}$ from all probes, similar to the compilation in \citet{Verde2019}, supplemented with measurements from the detailed compilation presented in \citet{diValentino2021} (their figure 1), and updated with the latest measurements. We add our estimate in blue, and we show the maximum contribution of the systematic error in black.}
\label{fig:H0}
\end{figure*}

In Figure \ref{fig:H0}, we compare this value with other probes from both Early-Universe and Late-Universe measurements. For the Early-Universe data, ``{\it Planck}'' refers to the {\it Planck} satellite's {\tt plik} likelihoods for the TT, TE and EE likelihood power spectra~\citep{Planck2018_para}. ``{\it Planck}+BAO'' refers to the combination of {\it Planck} CMB temperature and polarisation data, with the BAO measurements at different redshifts. The BAO measurements include the standard ruler measured from the Six-degree Field Galaxy Survey (6dF), the Sloan Digital Sky Survey Data Release 12 (SDSS-DR12) main galaxy sample, and the SDSS-IV DR16 eBOSS LRG sample; see~\cite{Alam2021}. ``ACT DR4+{\it WMAP}'' refers to the ACT DR4 data (TT, TE and EE spectra), together with the {\it WMAP} data (TT and TE) at larger scales than those measured by ACT~\citep{Aiola2020}. ``ACT DR6+{\it Planck} lensing'' refers to the combination of ACT DR6 lensing and {\it Planck} lensing power spectra, with BAO measurements and a prior from big bang nucleosynthesis (BBN, \citealt{Madhavacheril2024}). ``DES+BAO+BBN'' refers to the combination of the Dark Energy Survey Year 3 results with the BAO measurements and the prior from BBN ~\citep{Abbott2022-DES}. ``{\it Planck}+ACT+DESI BAO'' is a recent measurement, and was obtained from the Dark Energy Spectroscopic Instrument (DESI) in conjunction with CMB anisotropies from {\it Planck} and CMB lensing data from {\it Planck} and ACT~\citep{Adame2024}. The latest measurement is ``ACT DR6+{\it Planck}+DESI BAO" from \citet{Louis2025} and uses ACT DR6 power spectra, combined with CMB lensing from ACT and  {\it Planck}, and BAO data from DESI Y1. All these Early-Universe data sit on the lower side in the range of current $H_0$ measurements and prefer a value between $67 - 68$ km s$^{-1}$ Mpc$^{-1}$.

\subsubsection*{Late-Universe measurements}

For the Late-Universe data, ``SH0ES'' refers to the measurement of $H_{0}$ with the calibration from Hubble Space Telescope observations of Cepheid variables in the host galaxies of $42$ SNIa~\citep{Riess2022}. ``TRGB'' refers to the measurement of $H_{0}$ based on a calibration of the tip of the red giant branch (TRGB) applied to SNIa~\citep{Freedman2019}. ``TDCOSMO'' refers to the measurement of $H_{0}$ from the strong-lensing time delay method, by using the spatially resolved kinematics of the lensed galaxy in RXJ1131-1231 obtained from Keck Cosmic Web Imager spectroscopy. The recent study by~\cite{Shajib2023} incorporates all uncertainties, including the mass sheet degeneracy inherent to the lens mass profile and the line-of-sight effects, as well as those related to the mass–anisotropy degeneracy and projection effects. ``GW170817'' refers to the measurement of $H_{0}$ from the ``standard siren'' technique, by using the localised merger of a binary neutron-star system (GW170817) in the galaxy NGC 4993~\citep{Abbott2017-GW}. \citet{Hotokezaka2019} updated this constraint by combining the gravitational wave measurement with radio observations of the superluminal motion of the jet. ``MIRAS'' refers to the measurement of $H_{0}$ using Mira-type stars (highly evolving low-mass stars) in a SNIa host to calibrate the SNIa luminosity~\citep{Huang2020}. ``Cepheids+TRGB+SBF'' refers to the Carnegie Supernova Project~I and II by combining various distance calibrator measurements, including Cepheids, TRGB, and Surface Brightness Fluctuations (SBF), which are derived from the {\it B}-band~\citep{Uddin2023}. 

``MASERS'' refers to the $H_{0}$ measured by the Megamaser Cosmology Project \citep{Pesce2020} using geometric distance measurements to six Megamaser-hosting galaxies in the CMB rest frame. This value is dependent on the method used to map peculiar velocities. \citet{Schombert2020} use the baryonic Tully-Fisher Relation (``TFR") for 50 galaxies with accurate distances from Cepheids or the TRGB. \citet{deJaeger2020} use Type II supernovae (``SNII") as standardisable candles. Lastly, a Hubble constant determination is given in \citet{FernandezArenas2018}, that uses the relation between the integrated H$\beta$ line luminosity and the velocity dispersion of the ionized gas of HII galaxies and giant HII regions as a standard candle.

The Late-Universe measurements lie on the higher side in the range of current $H_0$ measurements. The SH0ES estimate, which has the smallest error, is in a clear $5\sigma$ discrepancy with the combined {\it Planck}+BAO results. We also indicate an ``Optimistic average" of 23 Hubble constant measurements based on Cepheids+SNIa, TRGB+SNIa, Miras+SNIa, Masers, Tully Fisher, Surface Brightness Fluctuations, SNII, Time-delay Lensing, Standard Sirens and $\gamma$-ray Attenuation, obtained by \citet{DiValentino2021_2}. We also show the ``Ultra-conservative" estimate by \citet{DiValentino2021_2} which was obtained removing the Cepheids+SNIa and the Time-Delay Lensing based measurements. These averaged estimates also emphasise the tension with the $\Lambda$CDM model. It is also important to note that the measurements from TDCOSMO~\citep{Shajib2023} and GW170817~\citep{Abbott2018} exhibit considerable uncertainties, making them consistent with Early-Universe estimates. Although our measurement, which pertains to a low-redshift regime, provides a much lower value for $H_0$, the large errors associated with our data allow for consistency with both Early- and Late-Universe measurements.

\section{Conclusions}
\label{summary}


In this study, we employ the cosmic chronometer (CC) method to measure the Hubble parameter at redshift $z=0.5$ using a sample of 53 brightest cluster galaxies (BCGs). The CC method provides cosmology-independent estimates of $H(z)$, as its calibration is not dependent on early-time physics or the traditional cosmic distance ladder. Furthermore, it is directly related to the value of the Hubble parameter at a specific redshift, rather than to the integral of $H(z)$, which is more commonly probed in angular diameter measurements (e.g., CMB and BAO).   

When all galaxies are considered, they form an envelope in the age--redshift plane. Very high number statistics are then needed to fill this envelope to probe the upper edge and measure the evolution of the oldest, most massive galaxies. The shallow slope of the upper edge of the envelope is a direct measurement of $H(z)$ (Eq.\ \ref{eqn:Hz}). Massive, passively evolving galaxies form a large part of this envelope, raising concerns that the mixing of different populations could affect the accuracy of the $H(z)$ measurements. Hence it is crucial to assess whether systematic uncertainties can be mitigated by selecting a more homogeneous sample of the oldest, most massive galaxies, brightest cluster galaxies (BCGs). Although previous CC studies have included small numbers of BCGs in their samples, this work focusses exclusively on BCGs from the BEAMS (BCG evolution with ACT, MeerKAT, and SALT) sample \citep{Hilton2021}. We used 53 BCGs between 0.3 $< z <$ 0.7 from BEAMS, without making any assumptions of the cosmological model. 

Using SALT optical spectra, we measure D4000$_{\rm n}$ for BEAMS BCGs, while accounting for all systematic uncertainties. By selecting a homogeneous sample of the most massive and oldest galaxies, we reduce the sensitivity to stellar metallicity, one of the largest sources of systematic error. Our new direct measurement of $H(z)$ at $z=0.5$ yields $72.1 \pm 33.9(\rm stat) \pm 7.3$(syst) km s$^{-1}$ Mpc$^{-1}$, and the projected value of $H_{0}$ is $54.6 \pm 25.7(\rm stat) \pm 5.5$(syst) km s$^{-1}$ Mpc$^{-1}$. The combined systematic uncertainty is 10\%. 

As seen in Figures \ref{fig:Hz} and \ref{fig:H0}, current CC data do not allow us to weigh in significantly on the Hubble tension debate, with our BCG-only measurement providing an uncertainty on $H_{0}$ of the order of 47\%. Currently it serves as a valuable direct measurement and independent complement to other probes in the ongoing effort to resolve this issue. While the use of BCGs successfully reduces systematic biases, the statistical error remains a major limitation of the current study, and we consider how this can potentially be improved in future.

To determine galaxy ages using full-spectrum fitting is quite stellar population model dependent, even if it is a relative age measurement, as the systematic uncertainties pose limitations (see also discussion in \citealt{Kjerrgren2023}). We therefore consider how larger BCG samples could quantitatively improve the constraints on $H_{0}$ using D4000$_{\rm n}$ measurements in particular. Our errors on the D4000$_{\rm n}$ measurements from the stacked spectra are small, of the order of a few per cent (in the worst case 6\%). However, because of the slow evolution of D4000$_{\rm n}$ for massive, passive galaxies at this redshift, 12 data points (from 53 galaxies) with these measurements still propagate to an uncertainty of 47\% on the slope, therefore on $H(z=0.5)$ and by extension on $H_{0}$. The error on the D4000$_{\rm n}$ measurements depends on sample size, redshift range (i.e., apparent brightness of the galaxies), and data quality (e.g., telescope size, spectral resolution), whereas the propagation to the uncertainty on slope depends on the expected evolution in that redshift range, number of bins chosen, etc. If we select random subsamples of our current data, and use the D4000n measurements from stacked spectra and the resulting slope to extrapolate and make rough predictions, we find that in order to have a D4000$_{\rm n}$ measurement with <1\% measurement error, we need to stack in excess of 50 BCGs per bin. For an uncertainty on the fitted slope to be $\sim$2\% or less, we need more than 50 such data points. Therefore, to achieve a $\sim$2\% statistical error, we need to observe more than 2500 BCGs in the same redshift range with the same data quality.

We can also use an existing CC study, which uses D4000n measurements of an extremely large sample of passive galaxies with M$_{*}$ > 10$^{11}$ M$_{\sun}$ (thus not only BCGs), as a guide. \citet{Moresco2016} used 130 000 galaxies from BOSS DR9 between 0.3 $< z <$ 0.5, which they divide in five redshift bins (see Section \ref{previousCC}). Even with these large statistics, it translates to an accuracy on $H(z)$ of 11 to 16\% per bin for their five measurements (light blue data points in Figure \ref{fig:Hz}), but once the entire sample is combined, and the systematic uncertainty contribution removed, a 1\% statistical uncertainty is reported. Of course using only BCGs decreases the number of galaxies with spectra available significantly.

At the moment, there is no dedicated survey to observe CCs (as for some other probes e.g., CMB, SNe, or BAO), but legacy data from other surveys, such as DESI \citep{DESICollaboration2016} will be most useful. Indeed, \citet{Wen2024} use the DESI Legacy Imaging Surveys data, supplemented with available spectroscopic redshifts, to identify 1.58 million clusters below $z < 1.5$. Due to the comprehensive nature of the sample selection of the DESI Luminous Red Galaxy (LRG) sample \citep{Zhou2023}, it is reasonable to expect that, upon completion, a significant number of the 8 million LRGs that will be observed with spectroscopy between 0.4 $< z <$ 1.0 will be BCGs. Equally exciting is the proposed Wide-field Spectroscopic Telescope (WST). In \citet{Mainieri2024}, the envisioned survey area, spectral resolution, and spectroscopic S/N achievable will provide a homogeneous way to observe the most massive, passive galaxy population across a very wide redshift range. Although no forecast is provided specifically for BCGs, their figure 61 predict that the WST is expected to reach an accuracy on $H(z)$ of the order of 5\% over the entire redshift range 0 < $z$ < 1.5 using CCs\footnote{This is presumably the total uncertainty, where some contribution from systematic errors were assumed.}. With such statistics, $H_{0}$ can be measured with statistical uncertainties competitive with the Cepheid+SNIa constraints on $H_{0}$. It is therefore also particularly important to have a good understanding of the systematic uncertainties, the limitations, and think of possible ways in which it can be further reduced.

\section*{Acknowledgements} 

We thank the anonymous reviewer for their constructive comments that improved the paper. This work is based on research supported in part by the National Research Foundation (NRF) of South Africa (NRF Grant Number: CPRR240414214079). Any opinion, finding, and conclusion or recommendation expressed in this material is that of the author(s), and the NRF does not accept any liability in this regard. YZM acknowledges the support from the NRF with Grant Numbers 150580, 159044, CHN22111069370 and ERC23040389081. All long-slit spectra observations reported in this paper were obtained with the South African Large Telescope (SALT) under programme numbers 2019-1-LSP-001 and 2022-2-MLT-003 (PI: Hilton). This research used Astropy,\footnote{\url{http://www.astropy.org}} a community-developed core Python package for Astronomy \citep{Astropy2013, Astropy2018}, as well as $\mathtt{PySALT}$ \citep{Crawford2010}, $\mathtt{PyLick}$ \citep{Borghi2022a}, $\mathtt{pwlf}$\ \citep{pwlf}, $\mathtt{pPXF}$\ \citep{Cappellari2017}, \texttt{RSSMOSPipeline} \citep{Hilton2018}, $\mathtt{specstack}$ \citep{Thomas2019}, and $\mathtt{emcee}$ \citep{emcee2019}.  

\section*{Data availability}

The data underlying this article will be shared on reasonable request with the corresponding author (SIL, \url{ilani.loubser@nwu.ac.za}). More information on BEAMS data can be obtained from \url{https://astro.ukzn.ac.za/~beams/}, the corresponding author, or the BEAMS PI and the corresponding author for the ACT DR5 catalogue (MH, \url{matt.hilton@wits.ac.za}). The posterior likelihood data for the slope fitting (Figure \ref{fig:D4000n}) and the $Planck$+BAO16 $H(z)$ evolution band (Figure \ref{fig:Hz}) can be requested from YZM (\url{mayinzhe@sun.ac.za}). 



\bibliographystyle{mnras}
\bibliography{References} 



\appendix
\section{BCGs}
\label{bcgs_table}

We list the 53 BCGs, making up the 12 stacked spectra, in Table \ref{BCGs}.

\begin{table*}
\caption{The BEAMS BCGs making up the 12 stacked spectra. The cluster $M_{\rm 500c}^{\rm Cal}$ masses are from ACT DR5 \citep{Hilton2021}, and given in $10^{14}$  M$_{\sun}$.}    
\label{BCGs}      
\centering                      
\begin{tabular}{c l r r c c} 
\hline
 & Cluster (ACT-CL) & RA & Dec & $z$ (BCG) & Cluster mass ($10^{14}$  M$_{\sun}$) \\
\hline 
1 &     J0257.7--2209  & 	44.43276418 & 	--22.16035996 & 0.323 & 8.095$\pm$1.747 \\
 & J0405.9--4915 & 	61.493307 & 	--49.25582133 & 0.325  & 5.083$\pm$1.168 \\
 & J0014.0+0227 & 	3.502633772 & 	2.452013035 &  0.335 & 2.833$\pm$0.699  \\
 & J0127.2+0020 & 	21.81862977 & 	0.3473382 & 0.337 & 5.319$\pm$1.128  \\
 \hline
2 & J2031.8--4037	 & 307.9706837 & 	--40.61865041 &  0.344 & 10.487$\pm$2.232 \\
 & J0040.8--4408 & 	10.20584859 & 	--44.13367336 &  0.349 & 10.294$\pm$2.184  \\
 & J2325.1--4111 & 	351.2944561 & 	--41.19841058 &  0.355 & 7.619$\pm$1.605 \\
 & J0044.4+0150	 & 11.11632092	 & 1.835349671 &  0.358 &  4.751$\pm$1.032 \\
 & J0008.1+0201	 & 2.046719373 & 	2.020848324 &  0.364 & 6.414$\pm$1.356  \\
 & J0219.9+0130	 & 34.98013351 & 	1.501869584 &  0.364 & 4.145$\pm$0.919 \\
  \hline
3   &  J0028.3+0317  & 	7.084700771	  & 3.284592172   & 0.365 & 2.661$\pm$0.685   \\
  & J0024.6+0001	  & 6.154076742  & 	0.029123745  &  0.373 & 2.858$\pm$0.714  \\
  & J0239.8--0134	  & 39.95910019  & 	--1.569957603  &  0.377 & 9.239$\pm$1.920  \\
  & J0034.5+0225	  & 8.631972037  & 	2.41857106  &  0.382 & 8.123$\pm$1.700 \\
  & J0159.2+0030	  & 29.821137  & 	0.504201  &  0.382 & 2.962$\pm$0.731  \\
  & J0135.8--2044	  & 23.97342003  & 	--20.73904798    &   0.385 & 5.239$\pm$1.169  \\
   \hline
4 &   J0024.0--0444  & 	6.024763221  & 	--4.741720347  & 0.386 & 3.204$\pm$0.773  \\
  & J0416.1--2404   & 	64.04181493	  & 24.07454196  & 	0.395 & 10.524$\pm$2.179 \\
  & J0159.8--0849  & 	29.95917689  & 	--8.830646336	  & 0.396 & 9.406$\pm$1.928  \\
  & J0013.2--4906  & 	3.323826863  & 	--49.1117207  & 	0.402 & 6.839$\pm$1.407  \\
  & J0256.5+0006  & 	44.13320347  & 	0.104227181  & 0.407 & 5.632$\pm$1.215  \\
   \hline
 5 &  J0411.2--4819 &  	62.80684134 &  	--48.32252503 &  0.418 & 8.162$\pm$1.696  \\
 &  J0438.3--5419 &  	69.5750404 &  	--54.32055217 &  0.421 & 10.655$\pm$2.184  \\
 &  J0358.9--2955 &  	59.72696185 &  	--29.93020283 &  0.434 & 10.983$\pm$2.259  \\
 &  J0154.5--0039 &  	28.63188107 &  	--65.0787357 & 0.438 & 3.222$\pm$0.745 \\
  \hline
6 &  J0154.4--0321 &  	28.61830018 &  	--3.358369973 &  0.441 & 7.241$\pm$1.494  \\
 &  J2233.3--5338 &  	338.3294013 &  	--53.64644674 &  0.443 & 5.777$\pm$1.238  \\
 &  J0125.7--0634 &  	21.42888265 &  	--6.568795869 &  0.450 & 3.031$\pm$0.773  \\
 &  J0140.0--0554 &  	25.00682015 &  	--5.915780905 &  0.450 & 8.202$\pm$1.669 \\
 &  J2116.1--0226 &  	319.0266337 &  	--2.444540183 &  0.458 & 7.059$\pm$1.523 \\
 &  J0008.5--0519 &  	2.136488145 &  	--5.324074043 &  0.461 & 3.757$\pm$0.835  \\
  \hline
7 &  J2332.8+0109 &  	353.2163469 &  	1.166042632 &  0.483 & 3.244$\pm$0.722  \\
 &  J0025.8--0419 &  	6.473018574 &  	--4.323882853 &  0.492 & 3.822$\pm$0.842  \\
 &  J0005.7--3751 &  	1.440075832 &  	--37.85932487 &  0.503 & 5.930$\pm$1.186  \\
  \hline
8 &  J0014.9--0057 &  	3.726310569 &  	--0.950387441 &  0.534 & 7.059$\pm$1.412  \\ 
 &  J0132.2--0940 &  	23.0608236 &  	--9.681740655 &  0.538 & 3.010$\pm$0.826  \\
 &  J0006.9--0041 &  	1.728695689 &  	--0.688267729 &  0.541 & 3.862$\pm$0.805  \\
 &  J2335.1--4544 &  	353.7818597 &  	--45.74167253 &  0.548 & 6.000$\pm$1.183  \\
  \hline
9 &  J0045.2--0152 &  	11.302053	 &  --1.875472 &  0.548 & 7.147$\pm$1.419  \\
 &  J0046.0--0359 &  	11.517096	 &  --3.985813 &  0.550 & 4.189$\pm$0.889  \\
 &  J0137.4--0827 &  	24.35531649 &  	--8.459752465 &  0.565 & 8.793$\pm$1.709  \\
 &  J2345.0--0112 &  	356.265092 &  	--1.209885 &  0.566 & 2.866$\pm$0.637  \\
  \hline
 10 & J0046.4--3912 & 	11.60215182 & 	--39.20113631	 & 0.592 & 7.886$\pm$1.539  \\
 & J0240.0+0115 & 	40.01341585 & 	1.264729555 & 0.601 & 5.221$\pm$1.037  \\
 & J0502.9--2902 & 	75.73231299 & 	--29.03983108	 & 0.610 & 7.541$\pm$1.503  \\
 & J0559.7--5249 & 	89.93337941 & 	--52.82435631	 & 0.611 & 5.460$\pm$1.111  \\
  \hline
 11 & J0352.9--5648 & 	58.240247 & 	--56.80302889 & 0.614 & 4.921$\pm$1.009  \\
 & J0006.1--0231 & 	1.530118 & 	--2.524906 & 0.618 & 3.785$\pm$0.784  \\
 & J0030.8+0102 & 	7.70482863 & 	1.035445135 & 0.622 & 2.912$\pm$0.676  \\
 & J0008.8+0117 & 	2.224788372 & 	1.283422211 & 	0.625 & 2.731$\pm$0.640  \\
  \hline
12 & J0106.1--0618 & 	16.5403171 & 	--6.31656202 & 0.648 & 4.628$\pm$0.949  \\
 & J0026.2+0120 & 	6.566833228 & 	1.337233948 & 0.650 & 5.367$\pm$1.062  \\ 
 & J0021.5--4902 & 	5.390749708 & 	--49.0388432 & 0.668 & 3.319$\pm$0.714  \\
\hline                                   
\end{tabular}
\end{table*}			

\section{pPXF fits}
\label{ppxf_fits}

\begin{figure*}
\captionsetup[subfloat]{farskip=-0pt,captionskip=-0pt}
\centering
\subfloat[Stack 1]{\includegraphics[scale=0.38]{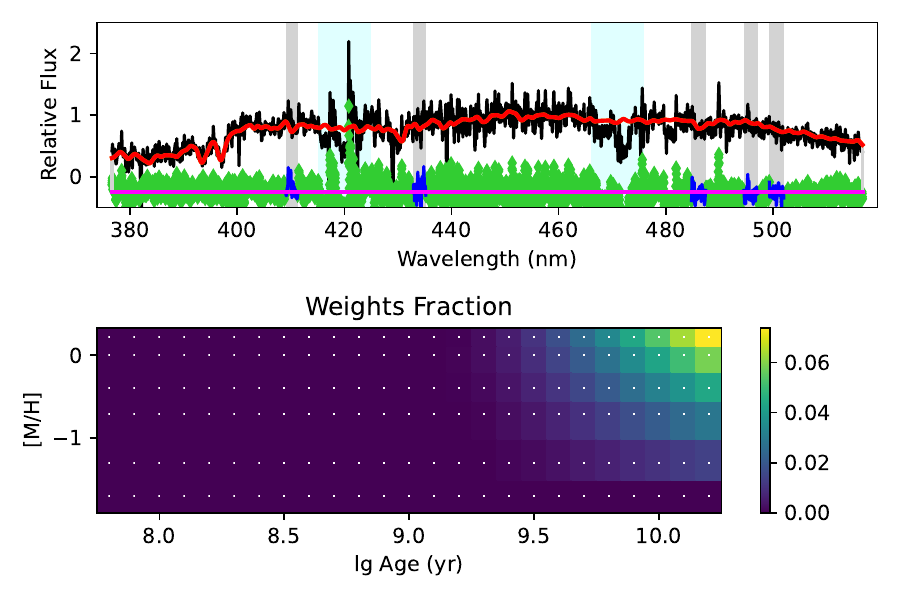}}
\subfloat[Stack 2]{\includegraphics[scale=0.38]{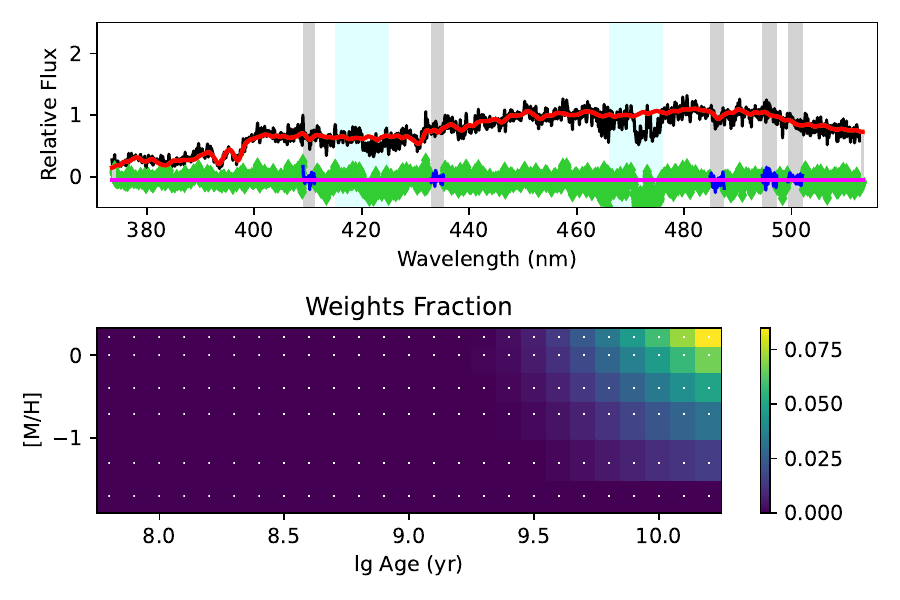}}
\subfloat[Stack 3]{\includegraphics[scale=0.38]{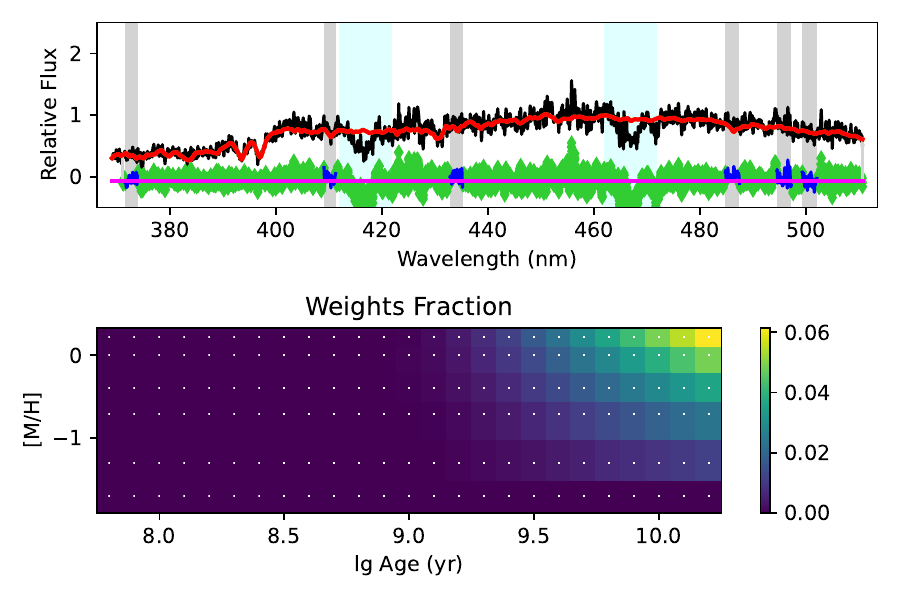}}\\
\subfloat[Stack 4]{\includegraphics[scale=0.38]{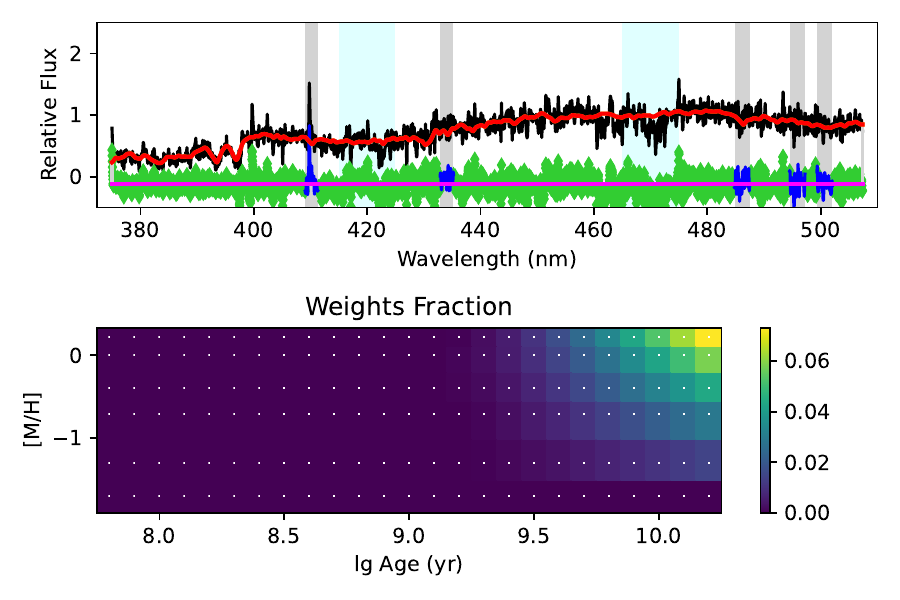}}
\subfloat[Stack 5]{\includegraphics[scale=0.38]{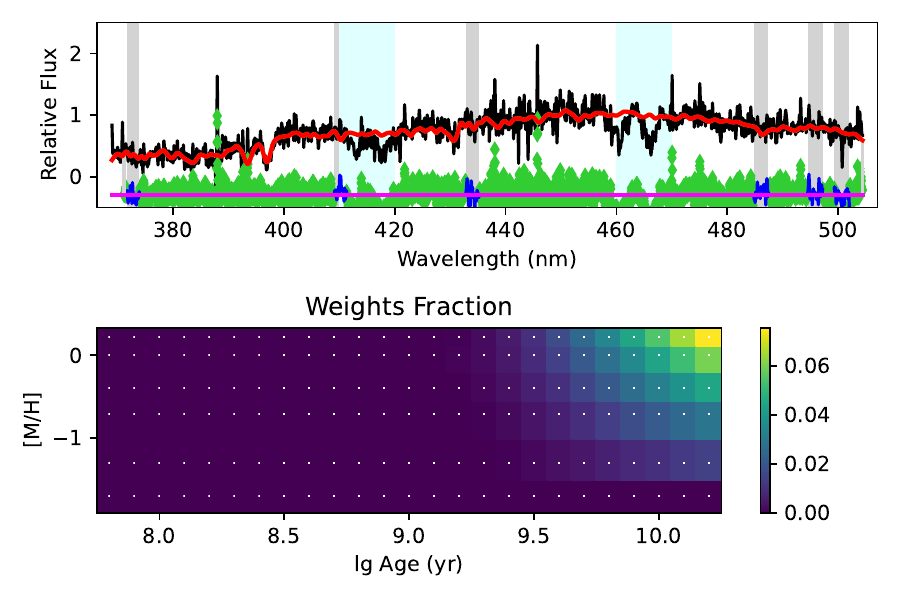}}
\subfloat[Stack 6]{\includegraphics[scale=0.38]{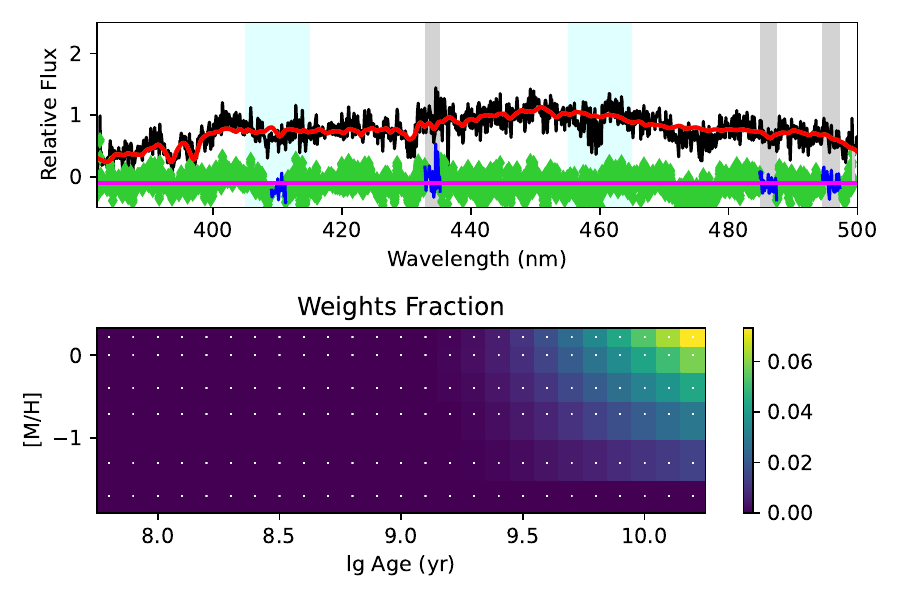}}\\
\subfloat[Stack 7]{\includegraphics[scale=0.38]{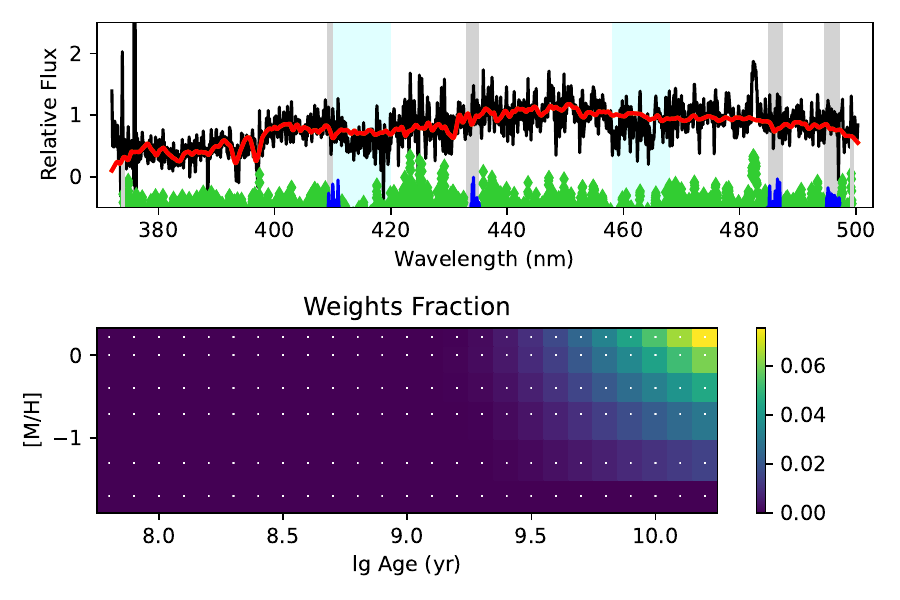}}
\subfloat[Stack 8]{\includegraphics[scale=0.38]{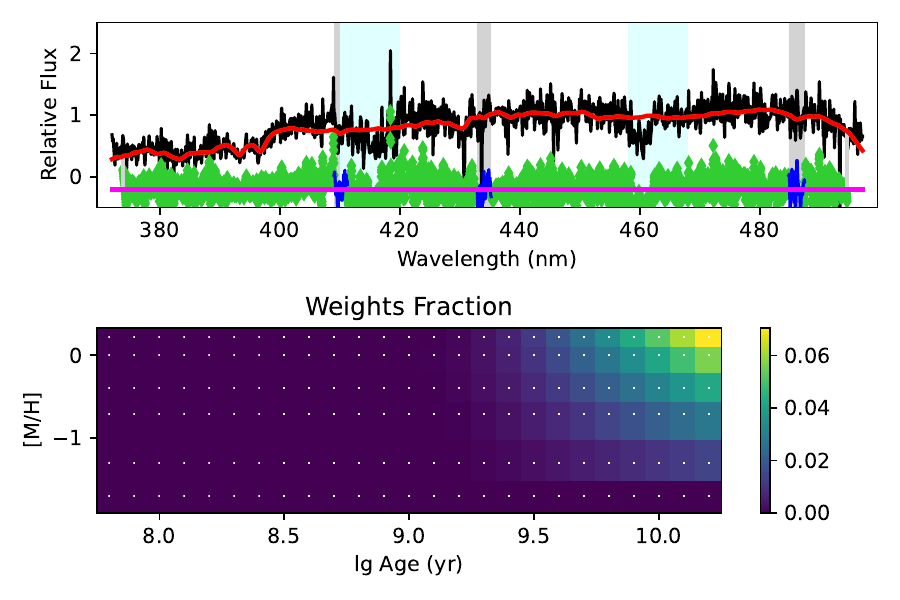}}
\subfloat[Stack 9]{\includegraphics[scale=0.38]{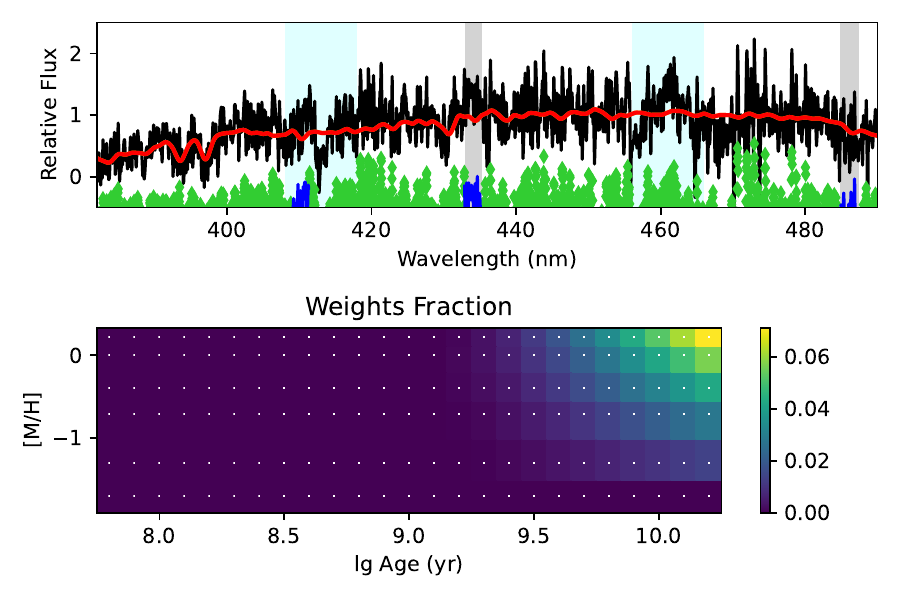}}\\
\subfloat[Stack 10]{\includegraphics[scale=0.38]{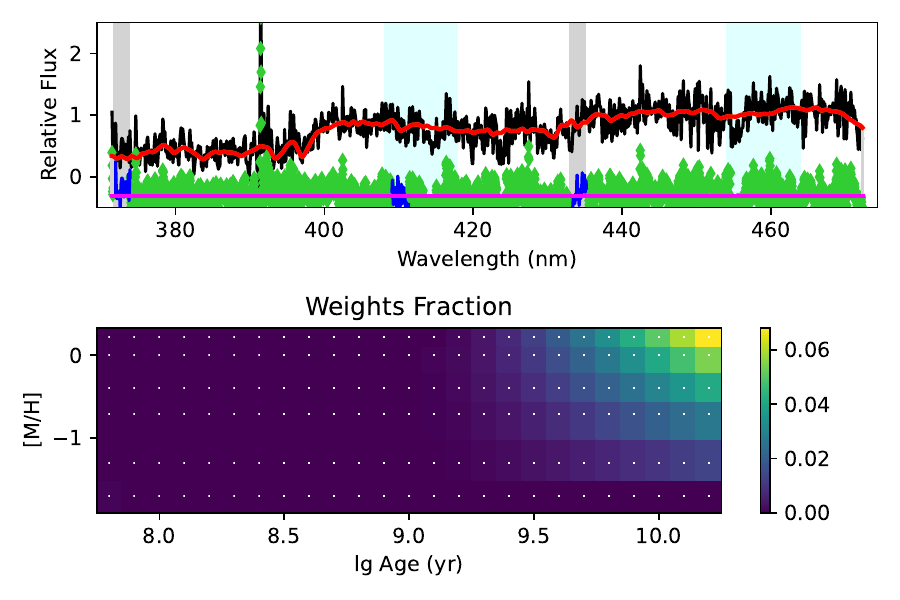}}
\subfloat[Stack 11]{\includegraphics[scale=0.38]{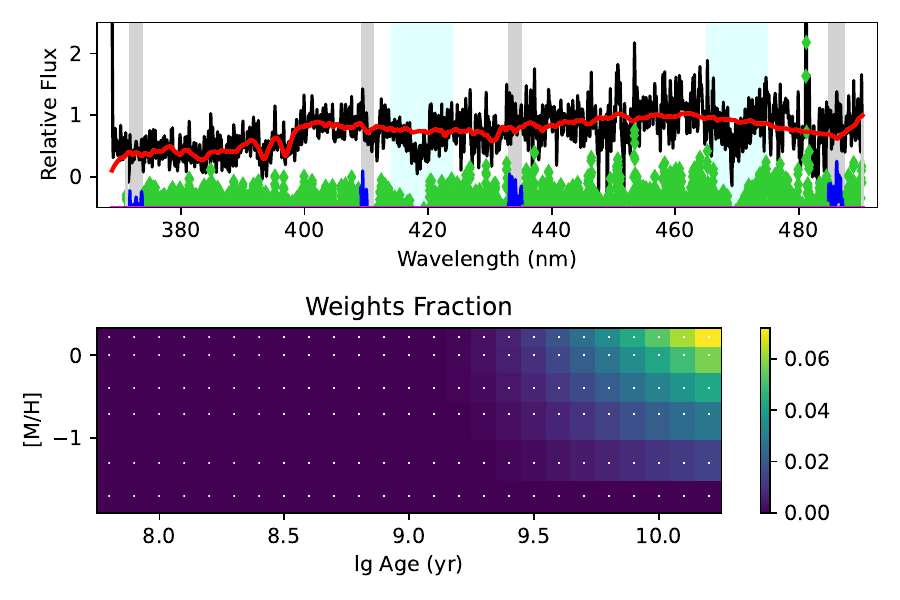}}
\subfloat[Stack 12]{\includegraphics[scale=0.38]{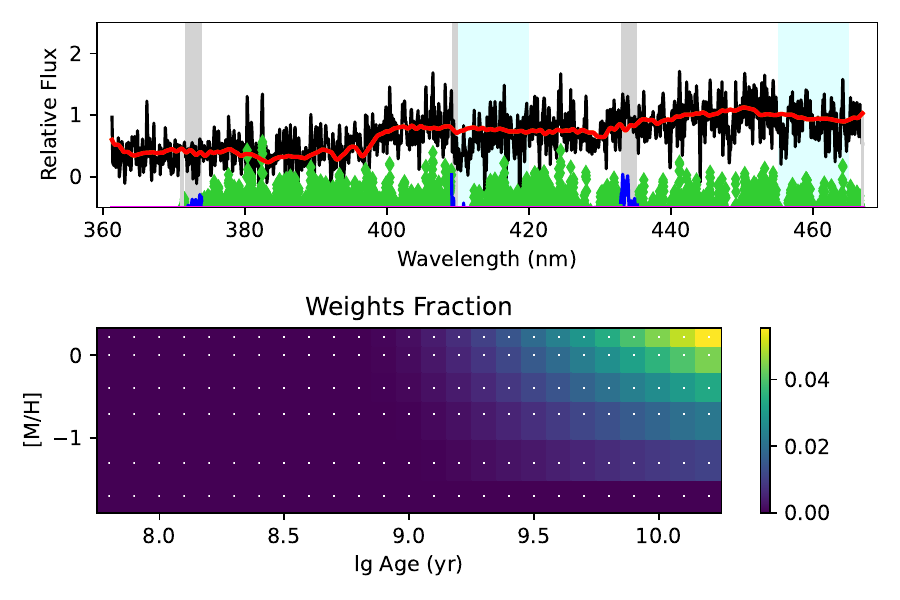}}
   \caption{Full-spectrum fitting of the stacks using \texttt{pPXF}. The observed stacked spectrum is shown in black, and the best-fitting combination of templates is shown in red. The grey bands indicate where emission lines will be found, if present. The light cyan bands indicate regions of the spectrum lost due to the SALT chip gaps. We also show the measured age and metallicity of the stellar populations which is, as expected, at the upper end of the range for the most massive, old, and metal-rich galaxies.}
\label{fig:stacked_spectra_ppxf}
\end{figure*}

We use \texttt{pPXF}, as described in Section \ref{fitting}, to fit the velocity dispersion, and at the same time we confirm the velocity is near zero, that there are no emission lines, and measure the stellar population age and metallicity. The fits are shown in Figure \ref{fig:stacked_spectra_ppxf}. The observed stacked spectrum is shown in black, and the best-fitting combination of templates is shown in red. The grey bands indicate where emission lines will be found, if present. The light cyan bands indicate regions of the spectrum lost as a result of the SALT chip gaps. We also show the measured age and metallicity of the stellar populations, even though they are all at the upper limits of the models.

\bsp 
\label{lastpage}
\end{document}